\documentclass{article}

\usepackage[english]{babel}

\usepackage[letterpaper,top=2cm,bottom=2cm,left=3cm,right=3cm,marginparwidth=1.75cm]{geometry}

\usepackage{comment}
\usepackage{amsmath}
\usepackage{amsfonts}
\usepackage{graphicx}
\usepackage[colorlinks=true, allcolors=blue]{hyperref}
\usepackage{algorithm}
\usepackage{algpseudocode}

\usepackage{caption}
\usepackage{subcaption}
\usepackage{multirow}
\usepackage{cite}
\usepackage{authblk}

\newcommand{\argmaxF}{\mathop{\mathrm{argmax}}\limits}   

\title{\LARGE \bf
Motion Prediction of Multi-agent systems with Multi-view clustering
}
\author{Anegi James$^{1}$ and Efstathios Bakolas$^{2}$}
\affil{University of Texas at Austin}

\date{}

\begin{document}
\maketitle

\begin{abstract}
This paper presents a method for future motion prediction of multi-agent systems by including group formation information and future intent. Formation of groups depends on a physics-based clustering method that follows the agglomerative hierarchical clustering algorithm. We identify clusters that incorporate the minimum cost-to-go function of a relevant optimal control problem as a metric for clustering between the groups among agents, where groups with similar associated costs are assumed to be likely to move together. The cost metric accounts for proximity to other agents as well as the intended goal of each agent. An unscented Kalman filter based approach is used to update the established clusters as well as add new clusters when new information is obtained. Our approach is verified through non-trivial numerical simulations implementing the proposed algorithm on different datasets pertaining to a variety of scenarios and agents.
\end{abstract}

\section{Introduction}

As the field of autonomous multi-agent systems continues to advance, motion prediction of such systems with varying levels of communication, similarity between system dynamics, operations and ability to reason is an essential concern. To achieve safe and reliable operation, these agents require accurate future predictions of the motion of the nearby agents in the surrounding environment, including other vehicles or humans and unexpected obstacles. Research towards predicting the motion of individual sub-groups such as road vehicles \cite{deo2018would}, drones \cite{2018Pengdrone}, UAVs \cite{sharma2009collision} and also robot-human interaction \cite{2013mainprice} scenarios is extensively explored.

In most settings which utilize an autonomous vehicle in motion, it is likely that the environment is populated by many different types of agents (e.g., pedestrians, bicyclists, vehicles), with their own distinct dynamics, motion and sensory capabilities. An agent might not have information or the ability to communicate with other agents in the domain but intentional grouping (such as groups of friends, cyclists travelling in groups) as well as unintentional grouping (such as those observed at pedestrian crossings or vehicles travelling at speeds restricted by traffic) is observed. 

\begin{figure}
    \centering
    \includegraphics[width=7cm, height=4cm]{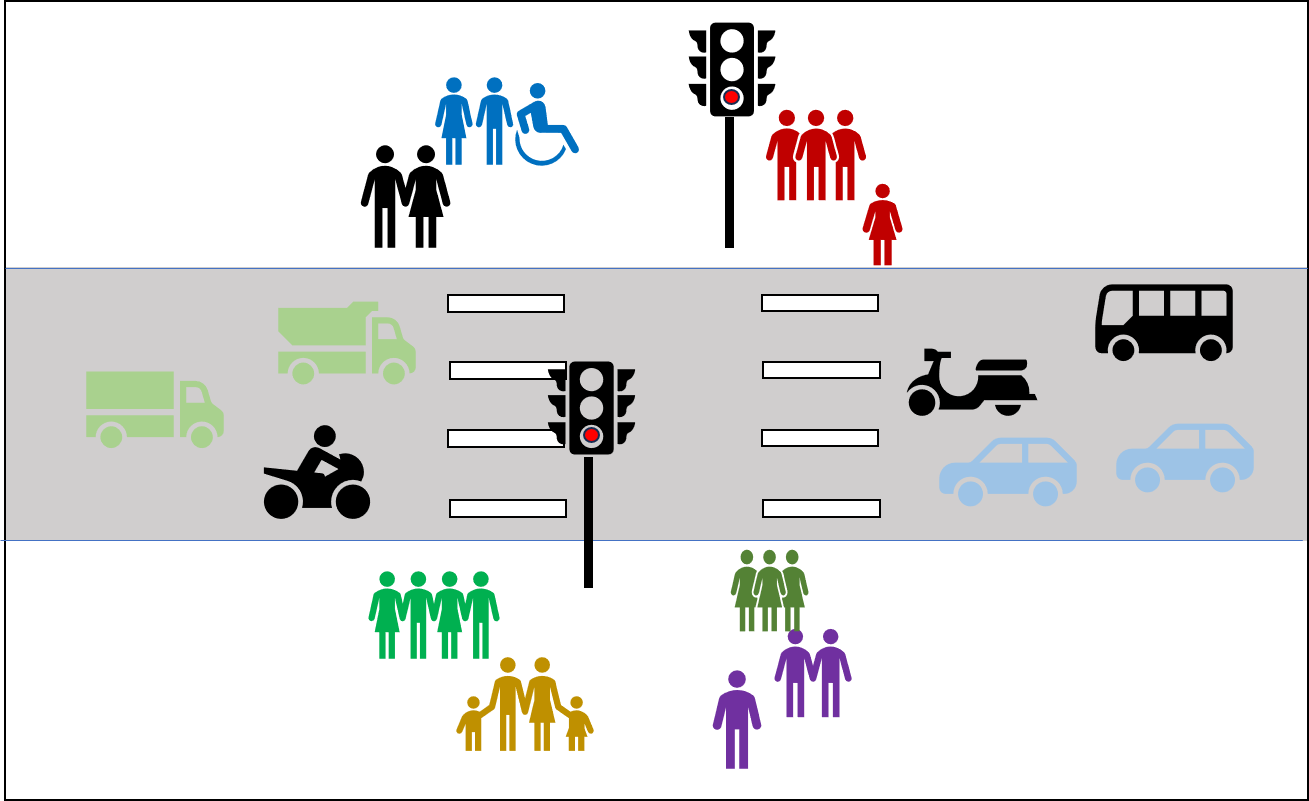}
    \caption{Multi-agent scenario at a crosswalk with both pedestrian and vehicular agents. An ego agent would have to predict the behaviour of various types of other agents in the environment for safe navigation. Agents highlighted with the same color indicate potential group formations based on their similar motion pattern, proximity and intent.}
    \label{fig:int1}
\end{figure}
In this work, this inherent grouping of agents in multi-agent scenarios is explored. Figure \ref{fig:int1} depicts one such scenario where we have multiple agents with differing dynamics and goals that must interact and navigate around each other. Various groups exist inherently for both the pedestrian and vehicular agents owing to personal relations and other connections such as proximity, environmental constraints and traffic patterns. This work aims to take advantage of the natural grouping present among agents to simplify the prediction process and provide truncated information for path planning. Accounting for these groups will reduce the computation time for prediction of the agent states and also identify denser regions of uncertainty that need to be avoided in the planning process.

\subsection{Literature Review}
Several authors \cite{luber2013multi, gorrini2015empirical, vizzari2013adaptive, wang2018detecting, cheng2014review} have proposed methods to detect the social grouping between pedestrians and identify or test several grouping models for urban crowds. Agent groups can be identified by analyzing predominant group features and patterns \cite{gorrini2015empirical} or by taking a data-driven approach. Data-focused approaches to finding groups/patterns in agent motion rely on data clustering models.  Clustering methods are most commonly used in trajectory planning to group trajectory pathlines into clusters which provide intuition for planning applications \cite{nguyen2021physics}, \cite{lee2007trajectory}.

A wide variety of clustering techniques exist to separate large-scale data based on some similarity measure between data points. Well-known clustering models relevant for the purpose of grouping agents include hierarchical agglomerative clustering \cite{shuming2002potential}, K-means, density based clustering \cite{kriegel2011density}, multi-view clustering \cite{hussain2014multi, yang2018multi, bickel2004multi} to name but a few. 
While these methods have different benefits in producing clusters with similar data points, they also have limitations depending on the application. For example, the K-means algorithm is widely used, but produces clusters of the same size and also requires the total number of clusters as an input parameter. 
Agglomerative clustering approaches circumvent the need for specifying the number of clusters a-priori, avoiding expensive computations for every clustering process.

Prediction of the agent groups may be completed through various estimation filters. Mixture filters permit the inclusion of the inter-dependent nature of sub-groups in a multi-agent system. In \cite{stordal2011bridging, kotecha2003gaussian, terejanu2011adaptive} variations of an adaptive Gaussian filter are proposed that include both the Kalman filter update and a method to re-sample the weights. The authors in \cite{guo2004multilevel} introduce a multi-level hierarchical Kalman filter method which samples from multiple levels to come to a consensus on the estimate. In \cite{wills2017bayesian}, authors propose a Bayesian Gaussian mixture filter that can reduce the mixture component number and prevent exponential growth. The authors of \cite{PGMI} and \cite{PGMII} propose two variations of a particle Gaussian mixture filter that propagates the state of the multi-agent system with particle filter methods and  Kalman measurement updates. The Unscented Kalman filter \cite{wan2000unscented} also provides a method for nonlinear estimation through approximation of the nonlinear dynamics with a minimum number of sigma points.

\cite{faubel2009split} introduces the split and merge unscented Kalman filter which is applied to nonlinear cases, removing mixture components with low probability and merging ones with similar means. Authors in \cite{khalkhali2019multi} and \cite{schulz2018multiple} adapt variations of a Kalman Filter for intention-based estimation of multi-agent systems.

\subsection{Contributions}
This paper contributes an improved method for efficient prediction of the evolution of the future Probability Density Function (PDF) and uncertain state trajectory of groups of agents which exploits their inherent tendency to form groups, building on our previous work \cite{james2022} which only utilizes a distance based metric to incorporate various motion features. A novel multi-view clustering method with a cost-based metric is incorporated into the clustering scheme along with the distance-based metric. An optimal control problem is defined which minimizes the control effort for an agent to exchange position and velocity with another agent in close proximity, resulting in an optimal cost-to-go relating the two agents. Additionally, a similar problem is solved to identify the optimal cost-to-go relating the agent and its known goal which provides a similarity criterion between agents based on their intent. The introduced cost-based clustering framework is able to identify agent groups based on features such as their relative distance and orientation while accounting for their dynamics which ensures identification of purposeful agent groups. Further, individual cluster member information is maintained along with the propagation of the cluster mean and covariance which can be extracted from the Gaussian mixture representation.

The state prediction method is extended to include the ability of the clusters to split and merge as well as update cluster membership with subsequent measurement updates providing updated state information. The validity of the proposed method has been tested on both pedestrian and vehicular agents using publicly available datasets namely: Argoverse 2 and Trajnet++.

\subsection{Outline}
Section II provides an overview of the preliminary information relevant to the agent dynamics and the Unscented Kalman Filter which is utilized for motion prediction. Section III details the clustering method to organize agents into clusters and cluster manipulation. Section IV formalizes the problem definition. Section V provides information on the simulations to test the method and the associated results.

\section{Preliminaries} 
\subsection{Notation} 
The vertical concatenation of two vectors $x \in \mathbb{R}^{n}$ and $y \in \mathbb{R}^{m}$ is denoted as $[x;y] \in \mathbb{R}^{n+m}$.
The Hausdorff Distance (HD) between two non-empty subsets $X,Y$ of a metric space is denoted by $d_{H}(X,Y)$, where
\begin{align*}
d_{H}(X,Y) = \max\{\sup_{x \in X} d(x,Y), \sup_{y \in Y} d(y,X)\},
\end{align*}
where $d(x,Y) = \inf_{y \in Y} \|x-y\|_2$, $d(y,X) = \inf_{x \in X} \|x-y\|_2$. 

Let $\rho_{\mathcal{N}}(x;\mu,P)$ represent the probability density function of a normal random variable $x$ with mean $\mu$ and covariance matrix $P = P ^{\mathrm{T}}>0$. 

\subsection{Agent Dynamics} \label{IIB}
Consider a group of $N$ mobile agents whose motion in the $2D$ plane is described by a double integrator kinematic model. In particular, the equations of motion of agent $i$, for $i \in \{1\dots N\}$, are given by 
\begin{equation} 
\ddot{p}^i_{x}(t) = u^i_{x}(t),~ \quad ~ \ddot{p}^i_{y} = u^i_{y}(t),
\end{equation}
where $p^i_{x}$, $p^i_{y}$ represent the coordinates of its position in the $2D$ plane (measured with respect to a given inertial frame), $v^i_{x}$, $v^i_{y}$ its velocity components and $u^i_{x}$, $u^i_{y}$ the acceleration components (the assumption here is that the agent can directly control its acceleration) at time $t\geq 0$. The equations of motion of agent $i$ can be represented in terms of a continuous-time state space model as follows:
\begin{align} \label{eqn:p1}
    \dot{x}^{i}(t) &= f(x^{i}(t),u^{i}(t))
\end{align}
where $x^{i} \in \mathbb{R}^{4}$ denotes the state of agent $i$ at time $t$ with $x^{i} = [p^{i}_{x} ; p^{i}_{y} ; v^{i}_{x} ; v^{i}_{y}]^{\mathrm{T}}$, $u^{i} \in \mathbb{R}^{2}$ denotes its control input with $u^{i} = [u_{x}^{i} ; u_{y}^{i}]$ and $f$ denotes its vector field with
\[
f(x^{i}, u^{i}) =  [v^i_{x}; v^i_{y}; u^i_{x}; u^i_{y} ]^{\mathrm{T}} = A x^{i} + B u^{i},\]
where, 
\begin{equation}\label{eq:matrices}
    A = \begin{bmatrix}
        \bold{0}_{2\times2} & \mathrm{I}_{2\times2}  \\
        \bold{0}_{2\times2} & \bold{0}_{2\times2} \\
    \end{bmatrix} 
    \qquad B = \begin{bmatrix}
        \bold{0}_{2\times2} \\
        \mathrm{I}_{2\times2} \\
    \end{bmatrix}
\end{equation}
We will also use the following notation for the state of agent $i$: $x^{i} = [p^{i}; \ v^{i}]$, where $p^{i} = [p^{i}_{x}; p^{i}_{y}]$ is the position vector of agent $i$ and $v^{i} = [v^{i}_{x};v^{i}_{y}]$ is its velocity vector.

We assume that each agent has a goal or desired state and can employ a steering law / controller that will steer it to the latter state. In particular, let us assume that agent $i$ is associated with the goal position, $p^{i}_{g} = [p^{i}_{x,g}; p^{i}_{y,g}]$, which tries to reach with zero velocity (soft landing) asymptotically (or exponentially) as $t \rightarrow \infty$ (obviously, the agent will ``practically'' reach the $p^{i}_{g}$ in finite time). The acceleration of the agent is composed of two acceleration components: 1) $u_{d}$ which represents the control action due to a proportional derivative controller and 2) $u_{f}$ which includes the interaction force between agents; in particular,
\begin{equation*}
    u^{i} = u^i_{d} + u^i_{f}
\end{equation*}
We assume that the acceleration of the agent $i$ for both $x$ and $y$ directions is conditioned on the proximity to the goal position in the 2D plane $p^{i}_{g}$ which should be reached with zero velocity, and is defined as follows:
\begin{align}\label{eq:PDcontrolsimple}
    u_{d}^{i}(x^i) &= K_{p}(p^{i} - p^{i}_{g}) + K_{v} v^{i} 
\end{align}

where $K_p, K_v \in \mathbb{R}$ are, respectively, the position and velocity gains which are assumed to have constant values for all the agents for simplicity.

The acceleration due to interaction between agents and agent groups is modelled using the social force model (SFM) which accounts for the attraction and repulsion between agents and their environment \cite{helbing1995social}. The interaction force, $F^{i}_{int}$, accounts for the repulsive force between agents in close proximity that mimics their tendency to avoid collisions with each other.  
Let the radius of an individual agent $i$ be given by $r_{i} >0$. The interaction force between two agents, $i$ and $j$ located at $p^{i}$ and $p^{j}$ respectively, is defined as:
\begin{equation} \label{eqn:sf_int}
    F^{i,j}_{int} =  A_{int} \exp \left( \dfrac{r_{i,j}-d_{i,j}}{B_{int}} \right)  n_{i,j}
\end{equation}
where $A_{int} > 0$ and $B_{int} > 0$ are constants, $d_{i,j} = \|p^{i} - p^{j}\|$ is the distance between the two agents, $r_{i,j}$ is the sum of the radius of both agents and $n_{i,j} = (p^{i} - p^{j})/d_{i,j}$ is the normalized vector pointing in the direction from agent $j$ to agent $i$. The overall interaction force on an agent $i$ is given by 
\begin{equation} \label{eqn:sf_i}
    F^{i}_{int} = \sum_{j = 1,i\neq j}^{N_{int}} F^{i,j}_{int},
\end{equation}
 where $N_{int}$ indicates the total number of other agents $i \neq j$ that are close to agent $i$ and included in its interaction zone such that $d_{i,j} \leq d_{tol}$. As $F^{i}_{int}$ depends on the distance between the agents, it is a function of the agent state $x^{i}$ and the dynamics in \eqref{eq:cloopDyn} have an additional non-linear component.
After closing the loop by using the feedback controllers given in \eqref{eq:PDcontrolsimple} and including the social interaction force in \eqref{eqn:sf_int} and \eqref{eqn:sf_i}, the equations of motion of agent $i$ are given by
\begin{equation} \label{eq:cloopDyn}
       \dot{x}^{i}(t)  = f^{i}_c(x^{i}(t)) =  A_{\mathrm{cl}} x^{i}(t) + g^i + F_{int}(x^{i}(t)),
\end{equation}
where 
\begin{align}
A_{\mathrm{cl}} = \begin{bmatrix}
        0 & 0 & 1 & 0 \\
        0 & 0 & 0 & 1 \\
        K_p & 0 & K_v & 0 \\
        0 & K_p & 0 & K_v 
    \end{bmatrix},~~~~g^i = -K_{p} \begin{bmatrix} p^{i}_{x,g} \\ p^{i}_{y,g} \end{bmatrix}.
\end{align}

\subsection{Clusters of agents} \label{IICluster}

Let us assume that the group of $N$ agents can be divided into a collection of $N_C \geq 1$ clusters (or sub-groups). Let $C_{J}$, with $J \in \{1, \dots, N_C\}$, denote the $J$-th cluster. In particular, $C_J$ consists of the indices of the agents that the cluster is composed of and let $| C_J |$ denote the cardinality of $C_J$. 
In this context, a cluster may comprise multiple agents moving as a group or an individual agent constituting a cluster group (in the latter case, $C_J$ corresponds to a singleton). 

To cluster $C_{J}$, we associate a state vector $\bar{x}_{J}$, where $\bar{x}_J = [\bar{p}_x^{J}; \bar{p}_{y}^{J}; \bar{v}_{x}^{J}; \bar{v}_{y}^{J}]$ which provides similar position and velocity information as with the individual agent case but for the cluster as a whole (each cluster is treated as a single ``representative'' agent). The cluster state corresponding to cluster $C_J$ is defined to be equal to the mean of the states of each individual agent in this cluster, that is, 
\begin{equation}
    \bar{x}_{J} =  (1/| C_J |) \sum_{i \in C_J} x^{i}.
\end{equation}
In the case of clusters with $| C_J |>1$ agents, the covariance matrix $\Sigma_{J}$ of this cluster is initially calculated using the statistical covariance, that is, 
\[
\Sigma_{J} = \frac{1}{| C_J |-1} \Sigma_{i \in C_J} (x^{i}-\bar{x}_J)(x^{i}-\bar{x}_J)^{\mathrm{T}}.
\]
For singleton clusters (that is, clusters  with $|C_J| = 1$), the covariance is set to a user-defined positive definite diagonal matrix, that is, $\Sigma_{J} = \mathrm{diag}(\sigma_{p},\sigma_{p}, \sigma_{v},\sigma_{v})$, where $\sigma_{p}, \sigma_{v} > 0$, are the variances of the position and velocity components, respectively (typically, both $\sigma_{p}$ and $\sigma_{v}$ should be taken to be sufficiently small numbers). Furthermore, the input of the representative agent of cluster $C_J$ is denoted by $\bar{u}_{J}$, where $\bar{u}_{J} = [\bar{u}_{x}^{J}; \bar{u}_y^{J}]$. The dynamic model used for the evolution of the state of the $J^{th}$ cluster in the 2D plane is also described by the dynamics given in \eqref{eqn:p1} but with the cluster mean positions and velocities, that is, 
\begin{equation} \label{cluster_dyn}
    \dot{ \bar{x}}_{J}(t) = f_{J}(\bar{x}_{J}(t), \bar{u}_{J}(t)),
\end{equation}
and the corresponding controllers (defined similarly to \eqref{eq:PDcontrolsimple})
\begin{equation} 
    u^J( \bar{x}_{J} ) = K_{p} (\bar{p}^{J} - p^{J}_{g})  +  K_{v} \bar{v}^{J} + F^{J}_{int}(\bar{x}_{J})
\end{equation}
where the gains $K_{p}$ and $K_{v}$ are the same as in the definition of the controller of the $i$th agent given in \eqref{eq:PDcontrolsimple} and the interaction force is defined as in \eqref{eqn:sf_i}. Consequently, the closed loop dynamics of the cluster $C_J$ is described by the same equation as in \eqref{eq:cloopDyn} after replacing $p^i_{g} = [p^{i}_{x,g}; p^{i}_{y,g}]$ with $p^J = [p^{J}_{x,g}; p^{J}_{y,g}]$.

\subsection{Unscented Kalman Filter} \label{IIUKF}
The Unscented Kalman filter (UKF) is a variation of the Kalman filter that approximates the estimated Gaussian PDF using sigma points. In this work, we adapt the formulation of the UKF for a multi-agent cluster setup. The sigma points associated with the UKF propagate and track the state and covariance of each individual cluster member $x^{i}$ in a given cluster $C_{J}$, instead of tracking only the cluster mean and covariance.

Next, we will briefly review the basic concepts and main steps of the UKF (the reader can find more details in \cite{sarkka2023bayesian}). Let the dynamics and measurement model of the agent be represented by the following discrete-time state space model:
\begin{align} \label{eqn:kf1}
    x_{k+1} &= f_{d}(x_{k}, \nu_{k}) \\
    z_{k+1} &= h_{d}(x_{k},\epsilon_{k}), 
\end{align}
where 

$x_{k}$ is the state of the single agent or the representative agent of the multi-agent cluster at time step $k$ (perhaps, a more precise notation for that state would have been $\bar{x}_{J,k}$, which we will not use here in order to keep the notation simple), $v_{k}$ is the process noise. Furthermore, $\epsilon_{k}$ is the measurement error associated with the measurement $z_{k}$. The measurement errors and process noise are assumed to be independently and identically distributed. 

The UKF is characterized by the selection of $L = 2n+1$ sigma points, $\chi_{i}, i \in \{1\dots L\}$, where $n$ is the dimension of the state space. 

For a cluster $C_J$ with $| C_J |$ number of members, let the associated cluster state vector be given by $\bar{x}_{J}$ and the state of the individual members be represented by $x^{j}$, where $j \in C$ is the index of the constituting members. The first sigma point is given by the cluster mean $\bar{x}_{J}$. For subsequent sigma points, the cluster member states $x^{j}$ represent the other $2n$ (or greater) cluster points as in \eqref{eqn:kfx2}:
\begin{equation}
\begin{aligned} \label{eqn:kfx2}
    \chi^{0}_{k} &= \bar{x}_{J} \\
    \chi^{i}_{k} &= x^{j}, \quad i = 1, \dots, 2n\\
\end{aligned}
\end{equation}
The weights associated with the states are given by $W^{m}$ and the covariance by $W^{c}$ and are calculated using an uneven distribution of weights. The weight associated with the $\chi^{0}$ sigma point is selected such that this point provides the majority of the contribution to the cluster state propagated by the UKF; in particular, $W^{m}_{0} = 0.5$. The weights of the other sigma points $\chi^{1:2n}$ are evenly distributed $W^{m}_{1:2n} = \frac{0.5}{2n}$. This weight distribution applies to the covariance case as well.

For singleton clusters consisting of only one member, the cluster mean $\bar{x}_{J}$ and the actual member state $x^{j}$ coincide and the usual formulation of the UKF is applied. The associated covariance with this single agent is given by $P_{x}$. In this case, we are required to generate the $2n+1$ sigma points following \eqref{eqn:kfx3}. The sigma points for singleton clusters $\chi_{k}^{i}, i=0\dots 2n$ with associated covariance $P = P^{\mathrm{T}} > 0$ are given by:
\begin{equation}
\begin{aligned} \label{eqn:kfx3}
    \chi^{0}_{k} &= x^{j} \\
    \chi^{i}_{k} &= \bar{x}^{j} + \sqrt{(n+\lambda)P} \quad i = 1 \dots n\\
    \chi^{i}_{k} &= \bar{x}^{j} - \sqrt{(n+\lambda)P} \quad i = n+1 \dots 2n \\ 
    \lambda &= \alpha^{2}(n+\kappa) - n,
\end{aligned}
\end{equation}
where $\alpha$, $\beta$ and $\kappa$ are user-defined algorithm parameters. In particular, $\alpha$ determines the distribution of the sigma points around the mean value and is set to $\alpha = 1$, $\kappa$ is a secondary scaling parameter, usually set to $0$ and $\beta = 2$ is the standard value for incorporating prior information given a Gaussian distribution \cite{wan2000unscented}. The weights are defined as follows:
\begin{equation}
\begin{aligned} \label{eqn:kf2}
    W_{0}^{m} &= \lambda/(\lambda + n) \\
    W_{0}^{c} &= \lambda/(n+\lambda) + (1 - \alpha^{2} + \beta) \\
    W_{i}^{m} &= W_{i}^{c} = 1/2(n+\lambda) \quad i=1\dots2n \\
\end{aligned}
\end{equation}
The weights associated with each of the sigma points $\chi_{k}$ are given by \eqref{eqn:kf2} for singleton clusters. $W^{m}$ indicate weights for the state propagation while $W^{c}$ terms represent the weighting factors for the covariance. 
The sigma points are propagated through the dynamic model $f_{d}$ as follows assuming process noise $\nu_{k}$:
\begin{equation} \label{eqn:kf3}
    \hat{\chi}^{i}_{k+1} = f_{d}(\chi^{i}_{k},\nu_{k})\quad i = 0\dots2n+1
\end{equation}
The mean and covariance for the agent's state through the prediction step are approximated as follows:
\begin{align} \label{eqn:kf4}
    \bar{y} &= \sum_{i=0}^{2n}W_{i}^{m}\hat{\chi}^{i}_{k+1} \nonumber \\ 
    P_{y} &= \sum_{i=0}^{2n}W_{i}^{c}(\hat{\chi}^{i}_{k+1}-\bar{y})(\hat{\chi}^{i}_{k+1} - \bar{y})^{\mathrm{T}}  +Q_{k}
\end{align}
The time updated sigma points $\hat{\chi}^{i}_{k+1}$ are propagated through the measurement model: 

\begin{equation}
    \begin{aligned}
        Z^{i} &= h_d(\hat{\chi}^{i}_{k+1}, \epsilon_{k})\\
        \bar{z} &= \sum_{i=0}^{2n}Z^{i}W_{i}^{m} \\
    \end{aligned}
\end{equation}
The measurement update leverages the sigma points to calculate the covariance matrices $P_{zz},P_{xz}, P_{k}$ for the update equation and the mean $x_{k+1}$ according to the following equations:
\begin{equation}
    \begin{aligned}
        P_{zz} &= \sum_{i=0}^{2n} W_{i}^{c}(Z^{i} - \bar{z})(Z^{i} - \bar{z})^{\mathrm{T}}\\
        P_{xz} &= \sum_{i=0}^{2n} W_{i}^{c}(\hat{\chi}^{i}_{k+1} - \bar{y})(Z^{i} - \bar{z})^{\mathrm{T}}\\
        P_{k} &= P_{y} - P_{xz}P_{xz}P_{zz}^{-1}\\
        x_{k+1} &= \bar{y} + P_{xz}P_{zz}^{-1}(z_{k} - \bar{z}) \\
    \end{aligned}
\end{equation}
where $x_{k+1}$ is the updated state given the measurement update $z_{k}$ at time step $k$.

\section {Methods}
\subsection{Similarity Metrics} \label{IIIA}
In order for two agents to be categorized as members of the same cluster, it is necessary for them to satisfy three conditions simultaneously: (1) they should be in close spatial proximity, (2) they should exhibit similar orientations and velocities in their respective motions, and (3) the agents should be moving towards goals in close relative proximity.
\subsubsection{Cost Distance}  \label{IIIACost}
An optimal control based metric is designed to implicitly account for the previously stated conditions for the similarity metric. This metric ensures that agents grouped in the same cluster are likely to move in similar directions in future time steps. 
The utilized metric corresponds to the weighted sum of the optimal cost functions of two separate but similar optimal control problems, which we will describe next.

The first problem concerns two agents, namely agent $i$ and agent $j$ which attain respectively states $x^{i} = [p^{i};v^{i}]$ and $x^{j} = [p^{j};v^{j}]$ at time $t\geq 0$ and governed by the agent dynamics defined in \eqref{eqn:p1}. The initial states of the two agents (at time $t=0$) are denoted by $x_{0}^{i}$ and $x_{0}^{j}$, where $x_0^{i} = [p_0^{i}; v_0^{i}]$ and $x_0^{j} = [p_0^{j}; v_0^{j}]$, respectively. The problem's objective is to find an optimal control input $u_{i}^*(\cdot)$ (1) that will transfer agent $i$ from the initial state $x^{i}(0) = [p^{i}_{0};v^{i}_{0}]$ to the final state $x^{i}(T_{f}) = [p^{j}_{0};v^{j}_{0}]$ (note that the terminal state of agent $i$ corresponds to the initial state $x_{0}^{j}$ of agent $j$) within the given time horizon $[0,T_{f}]$ and (2) will minimize the control effort:
\begin{equation} \label{eqn:pd1}
\begin{aligned}
    \min_{u_{i}} \quad J_1(u_{i}(\cdot)) & = \int_{0}^{T_{f}}\|u_{i}(t)\|^{2}\,\mathrm{d}t\\
\textrm{s.t.} \quad  \dot{x}^{i} & = f(x^i, u^i),~~~t\in[0,T_f]\\
\quad  x^{i}(0) & = x^{i}_{0} = [p^{i}_{0};v^{i}_{0}]\\
\quad  x^{i}(T_f) & = x^{j}_{0} = [p^{j}_{0};v^{j}_{0}]
\end{aligned}
\end{equation}
where $f$ is as defined in \eqref{eqn:p1}.
The solution of the optimal control problem given in \eqref{eqn:pd1} will provide us with a suitable metric for clustering agents that is neither distance-based nor explicitly dependent on its orientation or velocity (essentially, the optimal cost of the proposed problem can be viewed as the control effort required for agent $i$ to interchange its state with the state of agent $j$). 
As is shown in \cite{bakolas2021decentralized}, the optimal control input $u^{*}_i(t)$ that solves the optimal control problem \eqref{eqn:pd1}, where the velocity at the final time is substituted as $v(T_f) = v^{j}_{0}$, is given by
\begin{equation} 
    \begin{aligned} \label{eqn:costinput1}
    u_{i}^{*}(t) &= a + t b  \\
    b &= 1/T_{f}^{3}(12x^{i}_{0} - 12x^{j}_{0} + 6T_{f}v^{i}_{0} + 6T_{f}v^{j}_{0}) \\
    a &= -1/T_{f} (a T_{f}^2/2 + v^{i}_{0} - v^{j}_{0})
    \end{aligned}
\end{equation}
for all $t\in[0,T_f]$. The associated optimal cost, which is denoted by $\mathcal{V}_1(x_{0}^{i},x_{0}^{j})$, given the optimal control input $u_{i}^{*}(t)$ defined in \eqref{eqn:costinput1}, is calculated as: 
\begin{equation}
    \begin{aligned} \label{eqn:cost_optimal}
    \mathcal{V}_{1}(x^{i}_{0},x^{j}_{0}) &=  J_{1}(u_{i}^*(\cdot)) \\
    &= 1/2(T_{f}\|a\|^{2} + T_{f}^{2}a^{\mathrm{T}}b + 1/3 T_{f}^3\|b\|^{2}).
\end{aligned}  
\end{equation}
The cost function in \eqref{eqn:cost_optimal} serves as one of the measures of dissimilarity between agents belonging to different clusters. 

Similarly, the second component of the metric will be obtained by solving a second optimal control problem which seeks for the control input that will direct an agent to its (prescribed) goal destination while using minimum control effort. In particular, let agent $i$, whose motion is described by \eqref{eqn:p1} attain the state $x^{i} = [p^{i};v^{i}]$ at time $t\geq 0$ (its initial state at time $t=0$ is denoted by $x_{0}^{i}$, where $x_0^{i} = [p_0^{i}; v_0^{i}]$, as before), and let $x_{g}^i$, where $x_{g}^{i} = [p_{g}^{i};v_{g}^{i}]$, be its intended terminal (goal) state. Then, find a control input $u_i(\cdot)$ that (1) will transfer agent $i$ from its initial state $x^{i}_{0} = [p^{i}_{0};v^{i}_{0}]$, at time $t=0$, to the final state $x^{i}_{T_{f}} = [p^{i}_{g};v^{i}_{g}]$ at the final time $t= T_{f}$ for a given time horizon $T_{f}$, and (2) will minimize the control effort: 
\begin{equation} \label{eqn:cfp1}
\begin{aligned}
    \min_{u_{i}} \quad J_{2}(u_{i}(\cdot)) &= \int_{0}^{T_{f}}\|u_{i}(t)\|^{2}\\
\textrm{s.t.} \quad  \dot{x}^{i} & = f(x^i, u^i),~~~t\in[0,T_f]\\
x^{i}(0) & = x^i_0 = [p^{i}_{0};v^{i}_{0}]\\ 
x^{i}(T_f) & =x_{g}^i = [p^{i}_{g};v^{i}_{g}]\\
\end{aligned}
\end{equation}

The solution of the optimal control problem given in \eqref{eqn:cfp1} provides us with a suitable metric for clustering agents that also includes information about their intended goals, resulting in clusters that are potentially longer lasting.
The optimal control input $u^{*}_i$ that solves the optimal control problem \eqref{eqn:cfp1} is given by: 
\begin{equation} 
    \begin{aligned}  \label{eqn:costinput2}
    u_{i}^{*}(t) &= \alpha + t \beta  \\
    \beta &= -12/T_{f}^{3}(p^{i}_{g} - p^{i}_{0} + T_{f}v^{i}_{0}) - (6/T_{f}^{2}) v^{i}_{0} \\
    \alpha &= 6/T_{f}^{2}(p^{i}_{g} -p^{i}_{0} - T_{f} v^{i}_{0}) + (2/T_{f}) v^{i}_{0}
    \end{aligned}
\end{equation}
The associated optimal cost, which is denoted as $\mathcal{V}_2(x_{0}^{i},x_{g}^{i})$, is given by the right hand side of \eqref{eqn:cost_optimal} after replacing $x_{0}^{j}$ with $x_{g}^{i}$ and $a$, $b$ with $\alpha$, $\beta$ calculated in \eqref{eqn:costinput2}. 

The total cost distance is given by a weighted sum of the optimal cost functions associated with the solutions of the two optimal control problems we just described, that is,
\begin{equation}
    \label{eqn:sm1}
    \mathcal{V}(x_{0}^{i}, x_{0}^{j}; x_{g}^{i}) = \lambda_{1} \mathcal{V}_{1}(x_{0}^{i}, x_{0}^{j}) + \lambda_{2} \mathcal{V}_2(x_{0}^{i},x_{g}^{i}).
\end{equation}
The weighting factors $\lambda_{1}$, $\lambda_{2}$ are user-defined non-negative scalars and may be selected to reflect an interest in whether short term or long-term groupings are desired or depending on whether the agent intents are goal destinations or intermediate inferences on their projected path.

\subsubsection{Geometric Distance} \label{IIIAGD}
The closeness between individual agents is measured in terms of the Euclidean Distance (ED), $d_{i,j} = \|p^{i}-p^{j}\|_{2}$, while the Hausdorff distance measures the relative closeness of clusters. 
The Hausdorff Distance $d_{H}(C_{I},C_{J})$ measures the dissimilarity between collections of agents $C_{I} = (x_{1}, \dots, x_{N_{I}})$ and $C_{J} = (x_{N_{J}})$ comprising members of separate clusters. The Hausdorff distance is used when we are checking the inclusion of a singleton cluster ($C_{J}$) in another cluster ($C_{I}$). Both cost distance and geometric distance are considered for multi-view clustering of the agents into objective groups.

\subsection{Multi-view Clustering} \label{IIIBMVC}
The clustering of agents into different groups uses both cost and geometric distance methods for an agglomerative hierarchical clustering method that follows a bottom-up approach. Individual agents are first grouped into pairs and further pairs are grouped to form larger clusters. Clustering agents into pairs requires defining parameters and corresponding thresholds for comparison. The previous two sections detail the distance-based and physics-based clustering metrics.

Given a set of $N$ mobile agents identified with a point-set $X = \{x^{1}, \dots, x^{N}\}$ $\subset \mathbb{R}^{4}$ ($X$ is the set comprising the states of the $N$ agents), the process of estimating its future state is simplified by accounting for the presence of inherent groups in their interactions and movement. The identification of this grouping is performed using a multi-view clustering scheme that accounts for multiple factors characterizing the clusters. 

This section details the clustering process used to group each agent given their state predictions $x_{k+1}$ into an undefined number of clusters $N_{c}$. Let $\mathcal{C} = \{C_{1}, \dots ,C_{N_{c}}\}$ represent the set of clusters, where the $J$th member is represented as  $C_{J}$ and $\bar{x}_{J} = [\bar{p}_J;\bar{v}_{J}]$ defines the mean position and velocity of all the members in the cluster. In this work, the grouping between agents is considered to be influenced by their similarity metrics described in Section \ref{IIIA}.

Identifying groups within the multi-agent system requires finding a consensus between the multiple clustering views, i.e., the cost based and distance based view. Algorithm \ref{alg:al1} details the clustering scheme for the pairwise grouping of agents. Initially, every agent is assumed to belong to its own cluster. The minimum cost pair is calculated in lines 3-4 using \eqref{eqn:sm1} and the Euclidean distance for this pair is calculated in line 5. Agents are paired into a cluster, if there is consensus between their cost based metric and distance based metric, as calculated in lines 6-10.

\begin{algorithm}
\caption{Agent-Agent Grouping Algorithm}\label{alg:al1}
\textbf{Input} : $N > 0$, $x^{1},\dots, x^{N}$, $d_{tol}$\\
\textbf{Output} : $N_{c}>0$, \text{$C_{1}, \dots , C_{N_{c}}$}
\begin{algorithmic}[1]
\State \text{Define} $C_{1} = \{x^{1}\}$, \dots, $C_{N} = \{x^{N}\}$
    \For{\text{$i = 1:N$}}
        \State \text{Calculate cost for all other agents $j \neq i, j=1:N$}
        \State \text{Find min cost agent $x_{\min}$}
        \State \text{Find $\mathrm{ED}(x^{i},x^{j})$}
        \If {\text{ $\mathrm{ED}(x_{\min} < d_{tol}$)}}
        \State \text{Cluster $x^{i}, x_{\min}$}
        \State $N_{c} = N_{c} + 1$
        \State \text{Apply Algorithm 2 to cluster \& other agents $j$}
        \EndIf
    \EndFor
\State \text{Cluster agents into groups $C_{J}$}
\State \text{Calculate cluster means $\bar{x}_{J}$}

\end{algorithmic}
\end{algorithm}

\begin{figure}
     \centering
     \begin{subfigure}[b]{0.22\textwidth}
         \centering
         \includegraphics[width=\textwidth]{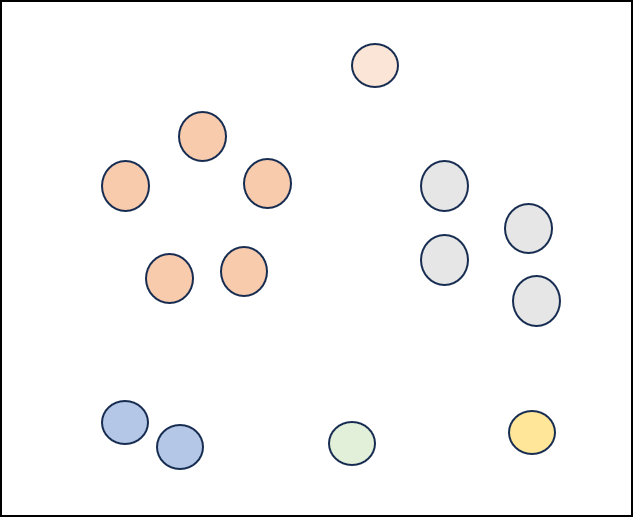}
         \caption{}
         \label{fig:ccg1}
     \end{subfigure}
     \hfill
     \begin{subfigure}[b]{0.22\textwidth}
         \centering
         \includegraphics[width=\textwidth]{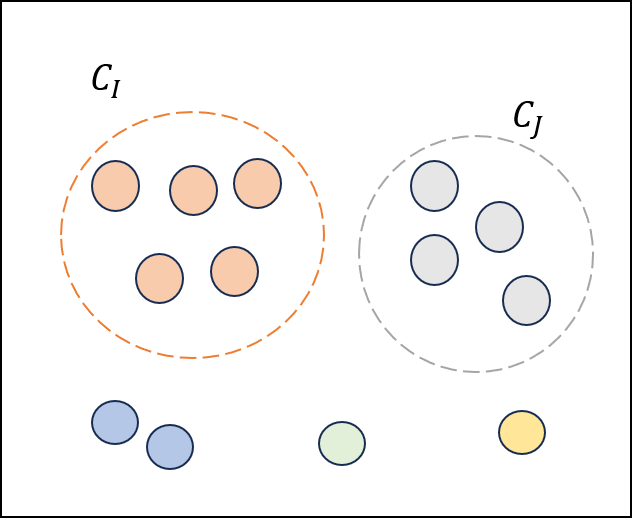}
         \caption{}
         \label{fig:ccg2}
     \end{subfigure}
     \hfill
     \begin{subfigure}[b]{0.22\textwidth}
         \centering
         \includegraphics[width=\textwidth]{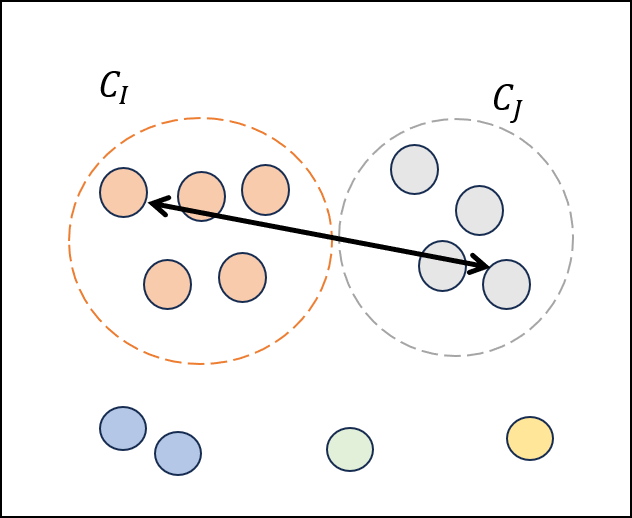}
         \caption{}
         \label{fig:ccg3}
     \end{subfigure}
     \hfill
     \begin{subfigure}[b]{0.22\textwidth}
         \centering
         \includegraphics[width=\textwidth]{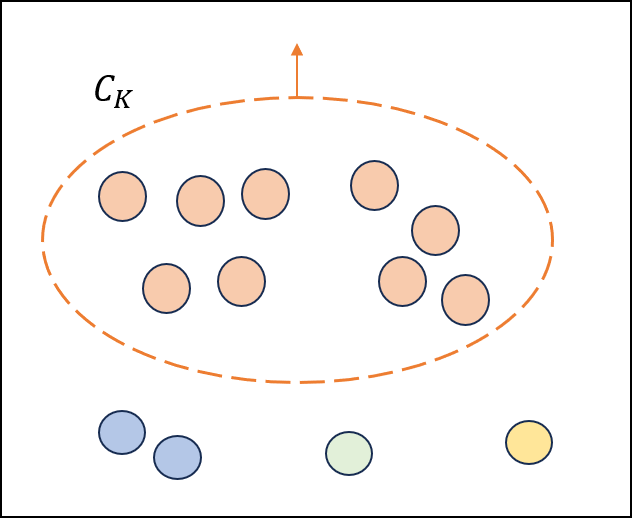}
         \caption{}
         \label{fig:ccg4}
     \end{subfigure}
        \caption{Example of complete linkage cluster-cluster grouping between two clusters $C_{I}$ and $C_{J}$. The shaded circles represent individual agents and the boundary indicates that they belong to one cluster. Complete linkage implies that the farthest agents of a group are similar. The clustering algorithm is applied to the agents connected by the solid black line in (c) to establish whether a complete linkage exists between the two clusters. The final cluster group is observed in (d).}
        \label{fig:cccgroup}
\end{figure}

Algorithm \ref{alg:al2} provides the description of the remainder of the clustering process for cluster-agent and cluster-cluster merging. Figure \ref{fig:cccgroup} depicts the complete linkage clustering method which accounts for the linkages between the farthest neighbors. Consider two clusters $C_{I}$ and $C_{J}$ with members ${i} \in C_{I}$ and ${j} \in C_{J}$. In line 2 of Algorithm \ref{alg:al2}, the agents with maximum distance between them, $i \in C_{I}$ and $j \in C_{J}$ are identified. Steps 3 and 4 calculate the cost associated with these agents using equation \eqref{eqn:cost_optimal} and find the cluster with the minimum associated cost. These clusters are merged, if their distances are within a predefined threshold $d_{tol}>0$ and the relative difference in cost is within a threshold $c_{\mathrm{tol}} > 0$. The tolerances may be chosen appropriately for the scale of the problem, for example, if the majority of the agents cover a smaller average distance over time, $d_{\mathrm{tol}}$ will have a smaller value. Any discrepancies in the capabilities of different agents in the same multi-agent scenario will be taken into consideration by the cost distance.

\begin{algorithm}
\caption{Cluster-Cluster Grouping Algorithm}\label{alg:al2}
\textbf{Input} : $C_{I}$, $C_{J}$, $d_{\mathrm{tol}}>0$, $c_{\mathrm{tol}} > 0$\\
\textbf{Output} : \text{$C_{1} \dots C_{M}$}
\begin{algorithmic}[1]
\State \text{Find $(x^{i},x^{j})$ =  $ \argmaxF_{x^{i} \in C_{I}, x^{j} \in C_{J}} \|x^{i} - x^{j}\|$}  
\State \text{Find optimal cost ($\mathcal{V}$) for agents $i \in C_{I}$ and $j \in C_{j}$}
\If  {\text{$ \mathcal{V}$($x^{i}$, $x^{j}$)} $<$ $c_{\mathrm{tol}}$}
    \State \text{Calculate Hausdorff distance ($d_H(C_{I},C_{J})$) between clusters}
    \State \text{Merge if $d_H \leq d_{\mathrm{tol}}$, $N_{c} \rightarrow N_{c} -1$}
\EndIf
\end{algorithmic}
\end{algorithm}

\subsection{Merge and Split Clusters} \label{IIIC}
Merging and splitting is part of the re-clustering process which takes place after the estimation step. Consider a set of clusters $\mathcal{C}$ with $N_{c}$ members where the $J^{th}$ cluster's state can be represented as $\bar{x}^{k}_{J}$ at some time $k$ with $J \in \{1,\dots,N_{c}\}$. Post the prediction and measurement update, the propagated set of clusters will go through a re-clustering process as described in Algorithms \ref{alg:al1} and \ref{alg:al2}. The resulting $J^{th}$ cluster state at time $k+1$ is given by $\bar{x}^{k+1}_{J}$ with $J \in \{1 \dots N_{q}\}$, where $N_q$ is the resulting number of clusters which has no dependence on $N_{c}$. This will account for whether a cluster remains the same, adds new members or removes members that are no longer relevant to the group. It is also the prior for the UKF in the next timestep.

\subsection{Insertion and Deletion of Clusters} \label{IIID}
New clusters are formed post prediction and estimation step, if there is an unassigned observation that cannot be associated with existing clusters or their corresponding members. Such an unassigned observation is added to the existing agent states $x^{i}$ and Algorithms \ref{alg:al1} and \ref{alg:al2} provide the new cluster set $\mathcal{C}^{k+1}$ with the updated information. In our case, the cluster/agent will be deleted due to the absence of an observation $z^{i}$ for a particular agent state $x^{i}$. However, the absence of a corresponding agent-observation pair does not necessarily indicate the irrelevance of that agent as the agent might have been missed by sensing systems for a number of reasons, including sensor fault or occlusion by the environment or other agents. 

\section{Problem Formulation} \label{IVPF}
\subsection{Problem Definition}

Consider a group of $N$ agents in motion with states denoted by $x^{i}$ and associated covariance $P^{i}$ as explained in Section \ref{IIB} and let the field of view be a bird's eye view of the domain. Each agent follows the dynamic model given in equation \eqref{eq:cloopDyn}. Time series data of state observations for all $N$ agents is available for time $t = t_{0}$. The associated probability density function $\rho_{\mathcal{N}}$ for this distribution is given by \eqref{eqn:pdn1} which provides information about the density of the field at each point $x$:
\begin{equation}\label{eqn:pdn1}
    \rho_\mathcal{N}(x;\mu^{i},P^{i}) = \frac{1}{\sqrt{2\pi |P^{i}|}} \mathrm{e}^{-\frac{1}{2}(x-\mu^{i})^{\mathrm{T}}{P^{i}}^{-1}(x-\mu^{i})}.
\end{equation}

We aim to predict the density evolution of the multi-agent system given an initial description of the multi-agent network's density. 

For a given set of clusters $\mathcal{C}$ , the PDF at the initial time is given by a convex combination of the PDFs of the clusters within the cluster set, that is, 
\begin{equation}
\label{eqn:pd2}
    \rho(t_{0},\hat{x}(t_{0})) = \sum_{i=1}^{N_{c}} w^{i}_{t_{0}} \rho_{\mathcal{N}}(\hat{x}(t_{0});\mu_{t_{0}}^{i},P_{t_{0}}^{i}),
\end{equation}
where the weights of the Gaussian mixture PDF are such that $w^{i}_{t_{0}} \geq 0$ and $\sum_{1}^{N_c} w^{i}_{t_{0}} =1$, $\forall i = 1,\dots,N_{c}$.
The main aim is to predict the future PDF of the clusters from the initial time to the final simulation time $t_f$ based on maximum likelihood estimation. The distribution of agents also varies with time and their association with each other might not remain constant over time. Identifying and accounting for these changes in the agent affiliations over time is also one of the main objectives.

The posterior PDF of each agent or cluster state is a Gaussian PDF. The posterior PDF distribution for the multi-agent system may be represented as a Gaussian mixture comprising the agent cluster position and velocity estimates at time $t_{f}$ using initially observed data which is given by
\begin{equation}
\label{eqn:pd3}
\rho(t_{f},\hat{x}(t_{f})) = \sum_{i=1}^{N_{c}} w^{i}_{t_{f}} \rho_\mathcal{N}(\hat{x}(t_{f});\mu_{t_{f}}^{i},P_{t_{f}}^{i})
\end{equation}
where the weights of the Gaussian mixture PDF are such that $w_{i} \geq 0$ and $\sum_{1}^{N_c} w_{i} =1$, $\forall i = 1,\dots,N_{c}$. We identify the PDF corresponding to the position of the agent in the domain for the entire time span of the simulation ($t \in [t_{0}, t_{f}]$). This provides a state estimate of every agent cluster at each time across the span and an overall density estimate of the occupied area in the domain. 

\subsection{Proposed Solution}
The UKF, described in Section \ref{IIUKF} is applied for state estimation of the multi-agent system. Provided initial state information at time $t_{0} = 0$, the agents are initially grouped into separate clusters based on Algorithm \ref{alg:al1} and \ref{alg:al2}, where each agent is either part of a cluster group or forms a singleton cluster by itself. The optimal value function of the optimal control problem \eqref{eqn:pd1} which is defined in \eqref{eqn:cost_optimal} provides a suitable metric for clustering agents that is neither distance-based nor explicitly dependent on its orientation or velocity. Algorithm \ref{alg:alg3} provides an overview of the process of computing  the prediction and update steps given a set of cluster states. Lines 2-3 of Algorithm \ref{alg:alg3}, set up the initial clusters as described in Section \ref{IIIBMVC}. Lines 5-7 update the cluster mean and covariances using the UKF described in Section \ref{IIUKF}. Any changes or updates to the clusters, including merging or deletion of clusters is indicated by the steps in lines 8-9.

\begin{algorithm}
\caption{Cluster motion prediction algorithm}\label{alg:alg3}
\textbf{Input} : $N_{c} \geq 0$, $N \geq 0$, $x_{0}^{1},\dots,x_{0}^{N}$, $T_f$\\
\textbf{Output} : \text{$\mathcal{N}(\mu,\Sigma) = \mathrm{UKF}(x^{1}_{T_{f}},\dots, x^{N}_{T_{f}})$}
\begin{algorithmic}[1]
\State $x^{i} \gets x_{0}^{i}$
\State \text{Cluster agents into groups $C_{J}$, $J \in \{1, \dots, N_{c} \}$}
\State \text{Calculate cluster mean $\bar{x}_{J}$}

\For{\text{$t = 1:T_{f}$}}
    \For{\text{$J \in \{1, \dots, N_c \}$}}
        \State \text{Run $\mathrm{UKF}$ on each cluster $C_{J}$}
    \EndFor
        \State \text{Add/remove clusters using observation update}
        \State \text{Re-cluster agent groupings based on latest update.}
\EndFor
\end{algorithmic}
\end{algorithm}

\section{RESULTS}
\subsection{Datasets}
The proposed algorithm for motion prediction of the clustered multi-agent system (that is, Algorithm 3 presented in Section \ref{IVPF} is tested with real-world datasets: Trajnet++ \cite{Kothari2020HumanTF} and Argoverse 2 \cite{Argoverse2}, \cite{TrustButVerify}. The Trajnet++ datasets provide time series data of the $x$, $y$ coordinates of the position of pedestrians, along with respective frame and scene information. Each scene is parsed through to obtain a dataseries with the required state of all agents on the scene. 

The Argoverse 2 Motion Forecasting dataset provides extensive position and velocity data collected from various cities separated into scenarios of 11$\text{s}$ segments with multiple types of agents such as vehicles, pedestrians, motorcycles and also static objects. 

\subsection{Simulation}
For each scenario/scene, we isolate the initial positions and velocities of all of the agents in that scenario which is passed on to Algorithm 1 and Algorithm 2 to group the agents into clusters using their initial states. Algorithm 3 is employed to predict the future state of the clusters of agents as well as of the individual cluster members. The algorithm output provides a future mean and covariance estimate for the predicted position and velocity of each agent and cluster at the final time $t_f$. Since the dataset for each sequence is complete, the algorithm is assumed to have access to observation data extracted from the original dataset at a lower frequency than the original. 
\subsubsection{Trajnet++}
\begin{figure}
    \centering
    \includegraphics[width = \linewidth]{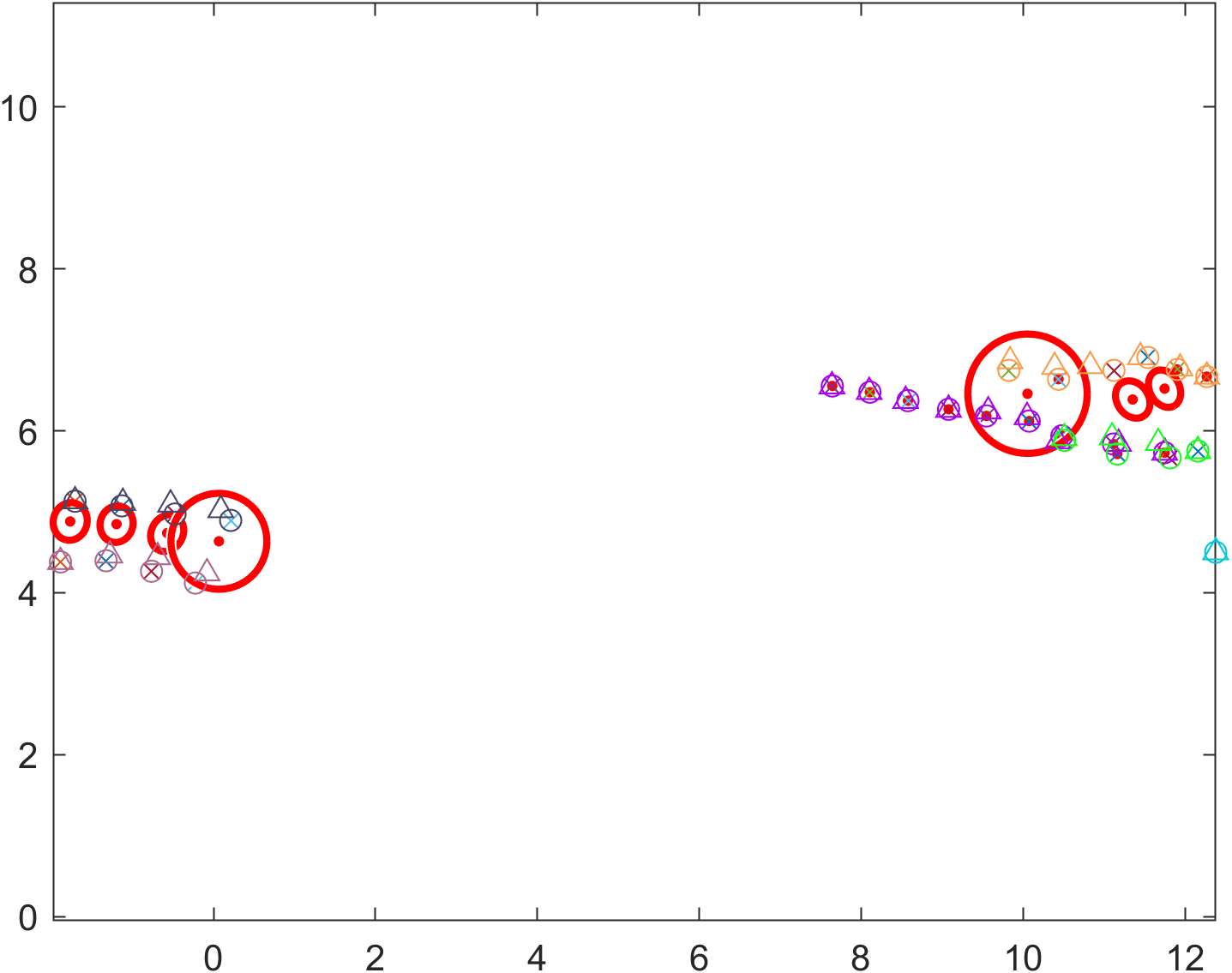}
    \caption{Prediction results on a scene from the Trajnet++ dataset. The circular markers represent the agent position predictions while the $'\times'$ markers represent the truth data for the position of the agent during the observed timeframe. The region enclosed by a red circle is the $3\sigma$ confidence ellipsoid around the cluster center. }
    \label{fig:t1}
\end{figure}

\begin{figure}
    \centering
    \includegraphics[width=\linewidth]{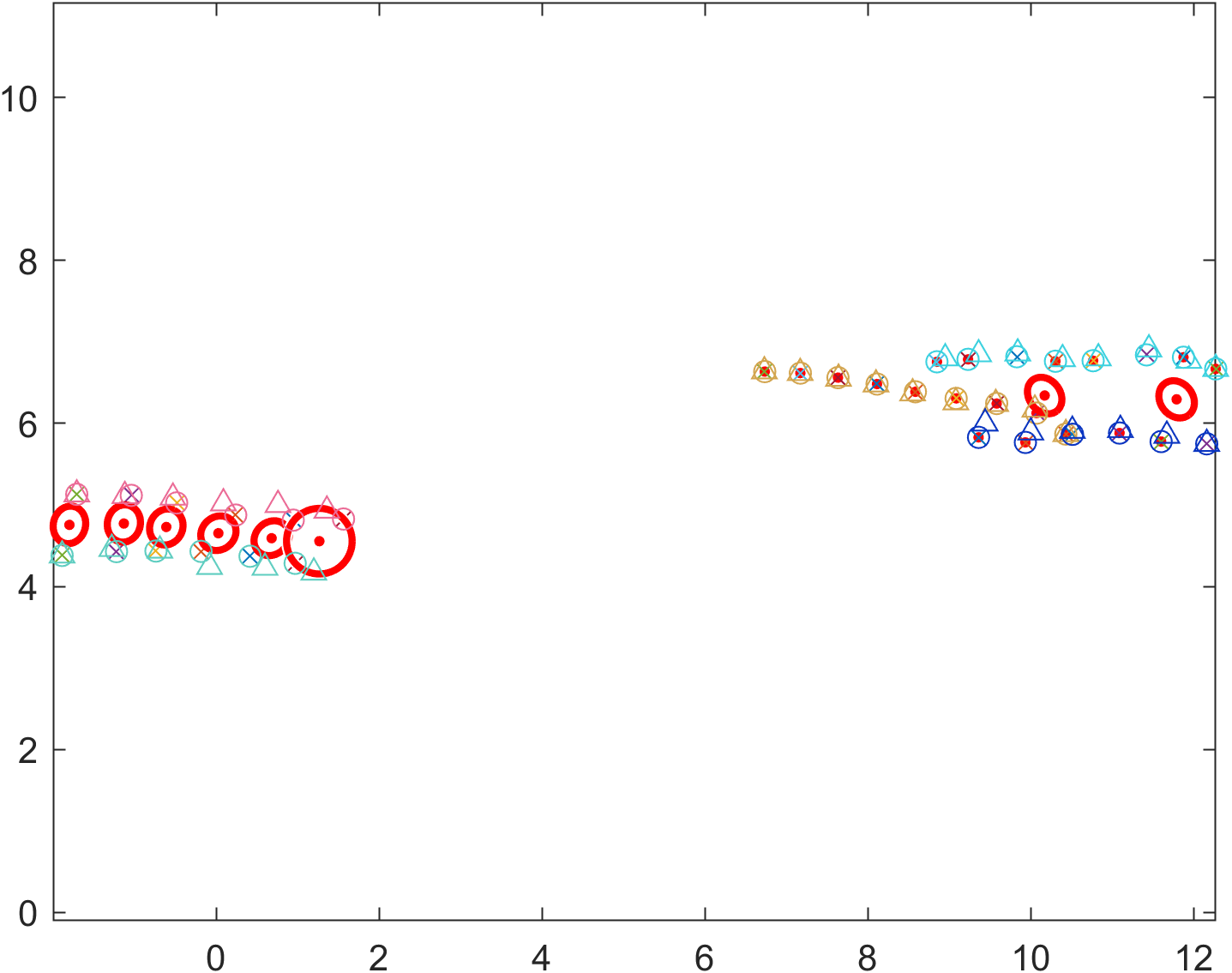}
    \caption{Prediction results on a pedestrian crossing scene from the Trajnet++ dataset. The circular markers represent the agent predictions while the $'\times'$ represent the truth data for the position of the agent during the observed time frame. The region enclosed by the red circle is the $3\sigma$ confidence ellipsoid around the cluster center. }
    \label{fig:t2}
\end{figure}

\begin{figure}
    \centering
    \includegraphics[width=\linewidth]{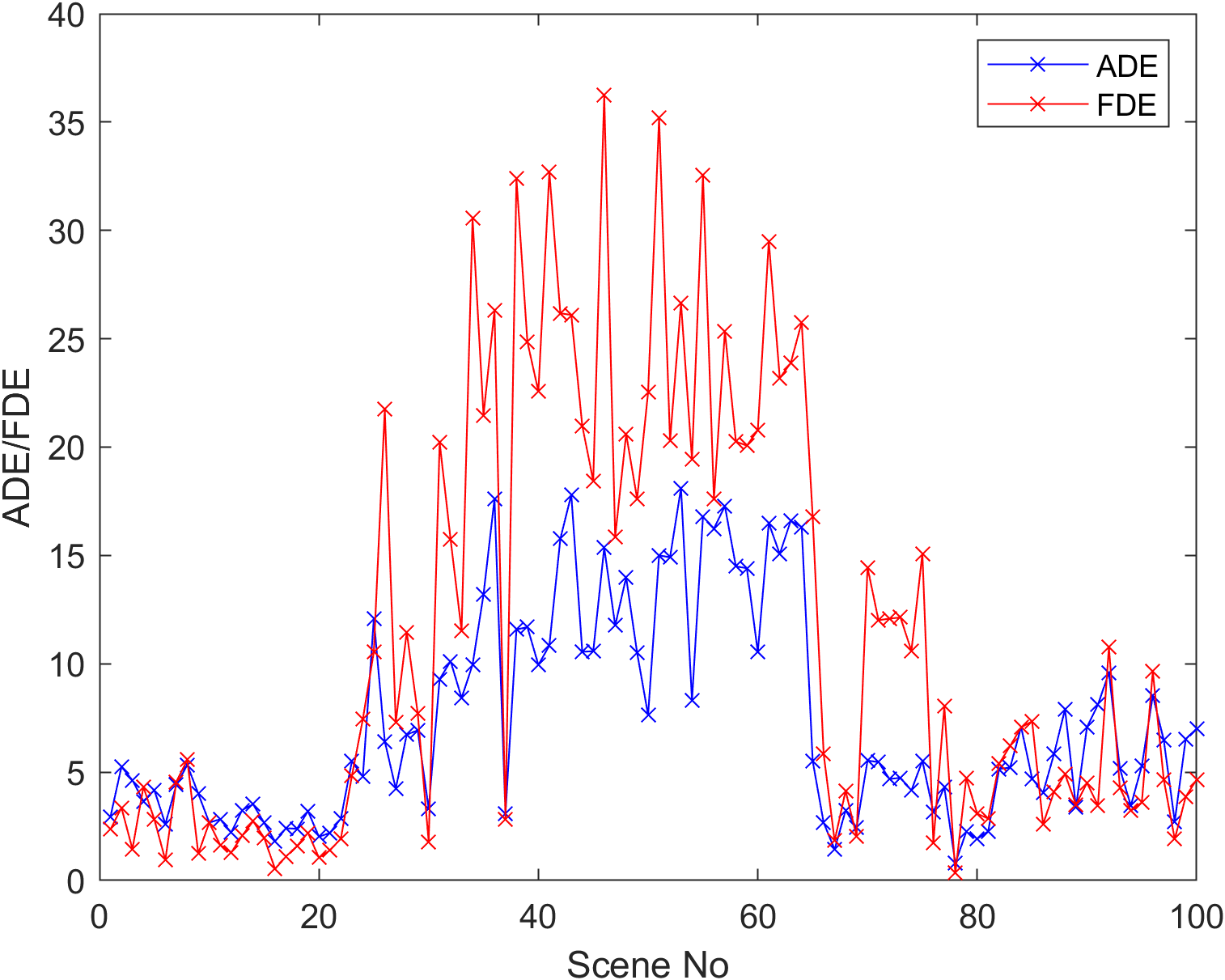}
    \caption{FDE and ADE calculated over several scenes for all the agents present on the scene. The average distance error in 'blue' is generally a lower value than the FDE in 'red' over the different scenarios.}
    \label{fig:t3}
\end{figure}

We will now present the clustering and prediction results based on data from the Trajnet++ which focuses on pedestrian agents. Figure \ref{fig:t1} and Figure \ref{fig:t2} represent the state mean predictions for two observed scenes of pedestrian data. The circular data points indicate predicted positions while the $'\times'$ markers represent the true agent data points. In these figures, the red circles represent the $3\sigma$ confidence ellipsoids around the cluster mean. The presence of smaller ellipses is due to the truth data sampling. Additionally, the predicted trajectory tracks the true trajectory with small errors. The errors can be attributed to the simplicity of the dynamic model not being able to capture the agent intents.
The average distance error (ADE) and final distance error (FDE) are calculated over a $100$ different scenes and their values are added up for all agents on the scene whose trajectory has been predicted. The FDE measures the difference in the final predicted position of the agent and the actual true position while the ADE averages the error over all timesteps. In Figure \ref{fig:t3}, the FDE has a larger value over all scenes in the range $10-80$ while the ADE averages under $5$ for the same scenes. While the final distance error is significant, the lower values of the ADE imply that the algorithm is able to predict an overall trajectory that matches the true trajectory for each of the observed agents on the scene.


\subsubsection{Comparison}
The clustering algorithm given by Algorithm 1 and 2 is compared against another standard clustering algorithm that uses only a geometric distance based criterion. Assume CD refers to the clustering algorithm from Section III and ED to Density Based Spatial Clustering, a Euclidean Distance based clustering algorithm. Table \ref{tab:table1} compares the result of both algorithms with several different scenes by evaluating the FDE and ADE as well as time taken for the simulation. Each scene is obtained from the Trajnet++ dataset and is less than $2$s in length. From the table, we observe that for the same tolerance parameters, our algorithm performs better than a solely distance-based algorithm that does not take into account the other features of the agent motion.
\begin{table}[]
    \centering
    \begin{tabular}{|c|c|c|c|c|} 
    \hline
    & Type & Time & FDE & ADE \\ 
    \hline
         \multirow{2}{*}{Scene 1}&ED & 0.0477 & 1.3051 & 1.2431\\
    &CD & 0.0067 & 0.633 & 1.1335\\  \hline
        \multirow{2}{*}{Scene 2}&ED & 0.011 & 6.9542 & 2.2399\\   
    &CD & 0.0048 & 3.1392 & 1.0674\\ \hline
    \multirow{2}{*}{Scene 3}&ED & 0.0094 & 23.7522 & 6.035\\ 
    &CD & 0.0064 & 0.0336 & 1.123\\  \hline
    \end{tabular}
    \caption{Comparison between ED and CD clustering schemes for three different scenes}
    \label{tab:table1}
\end{table}

\begin{figure}
     \centering
     \begin{subfigure}[b]{0.23\textwidth}
         \centering
         \includegraphics[width=\textwidth]{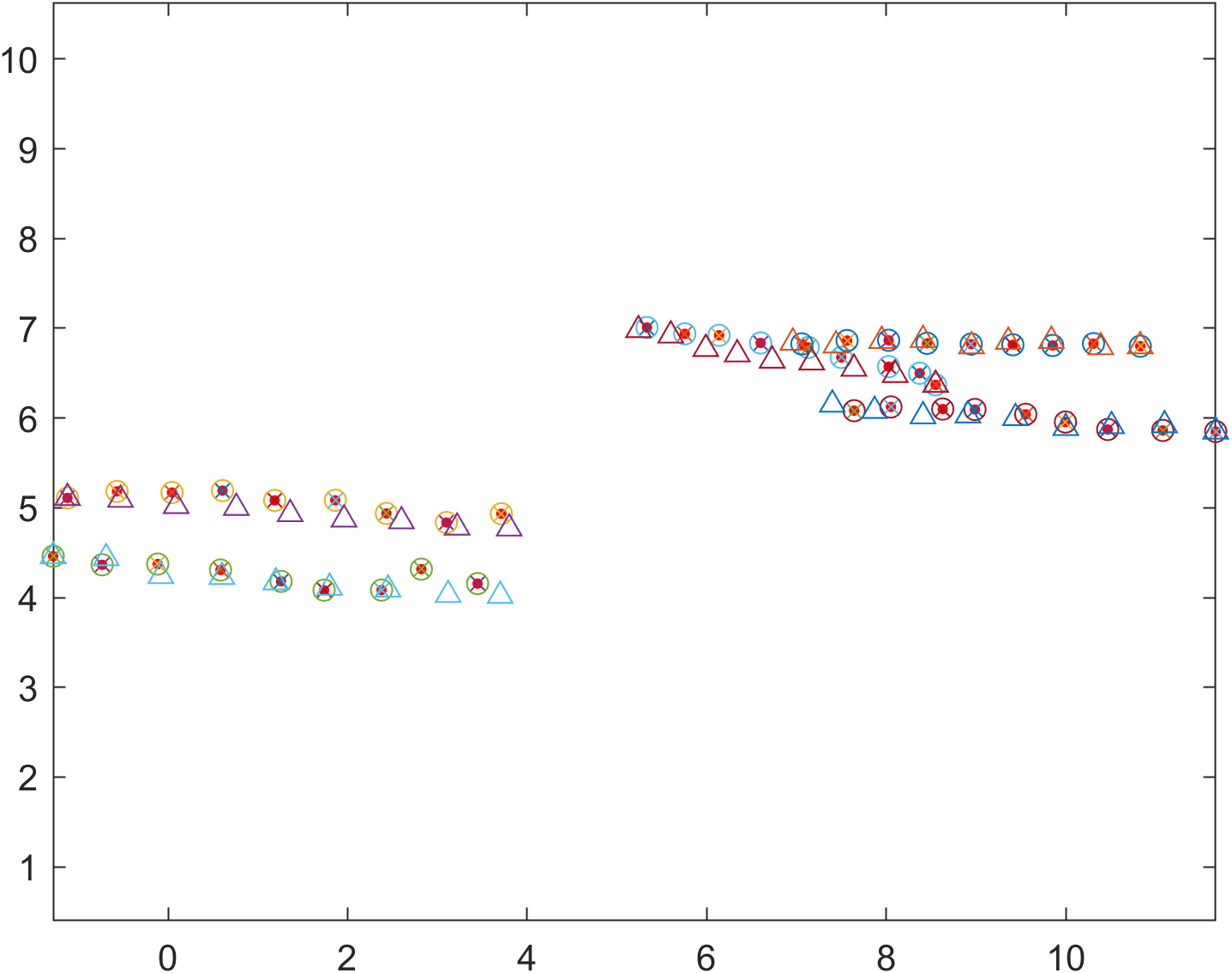}
         \caption{$\lambda_{1} = 0, \lambda_{2} = 1$}
         \label{fig:t4c1}
     \end{subfigure}
     \hfill
     \begin{subfigure}[b]{0.23\textwidth}
         \centering
         \includegraphics[width=\textwidth]{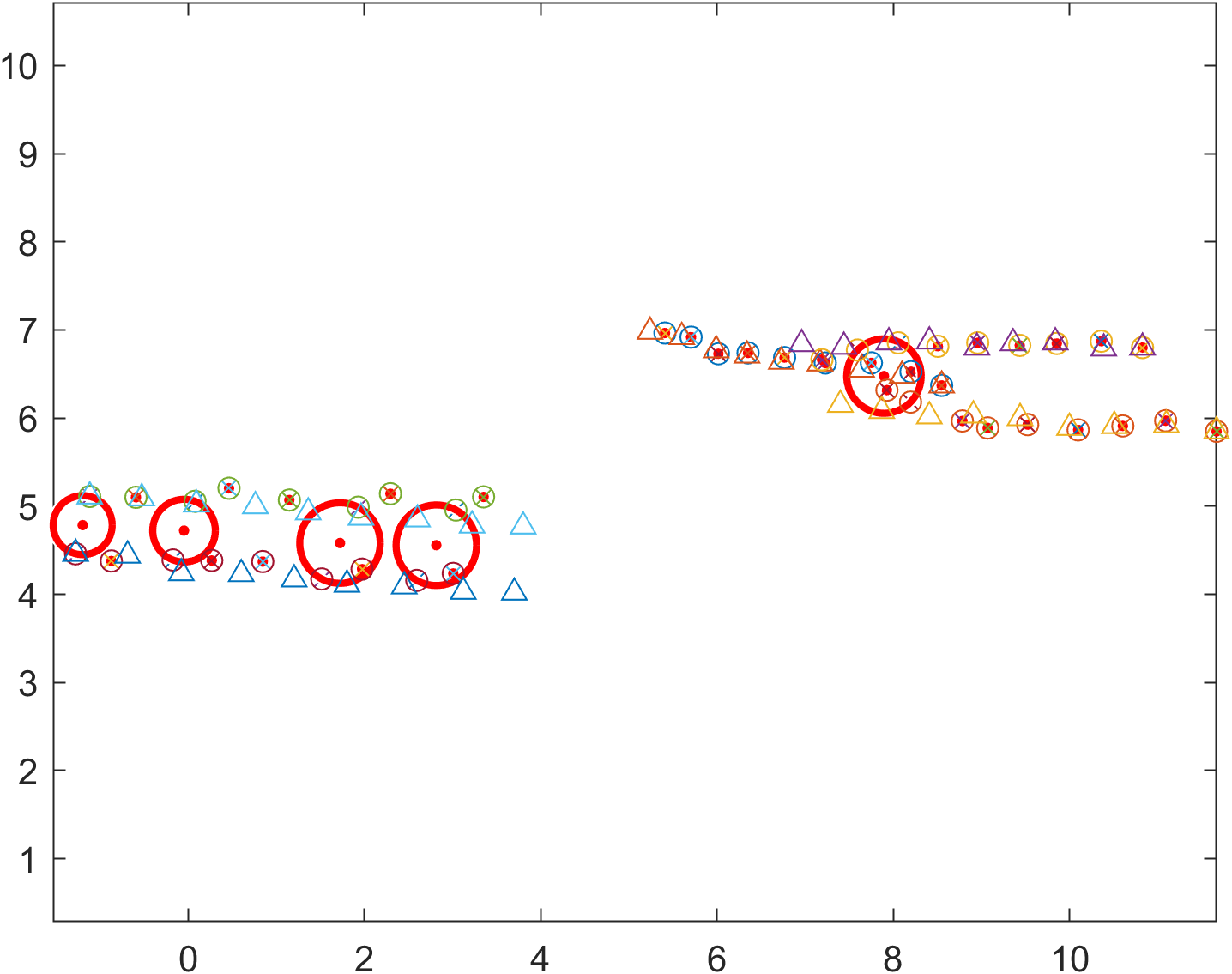}
         \caption{$\lambda_{1} = 0.5, \lambda_{2} = 0.5$}
         \label{fig:t4c3}
     \end{subfigure}
     \hfill
     \begin{subfigure}[b]{0.23\textwidth}
         \centering
         \includegraphics[width=\textwidth]{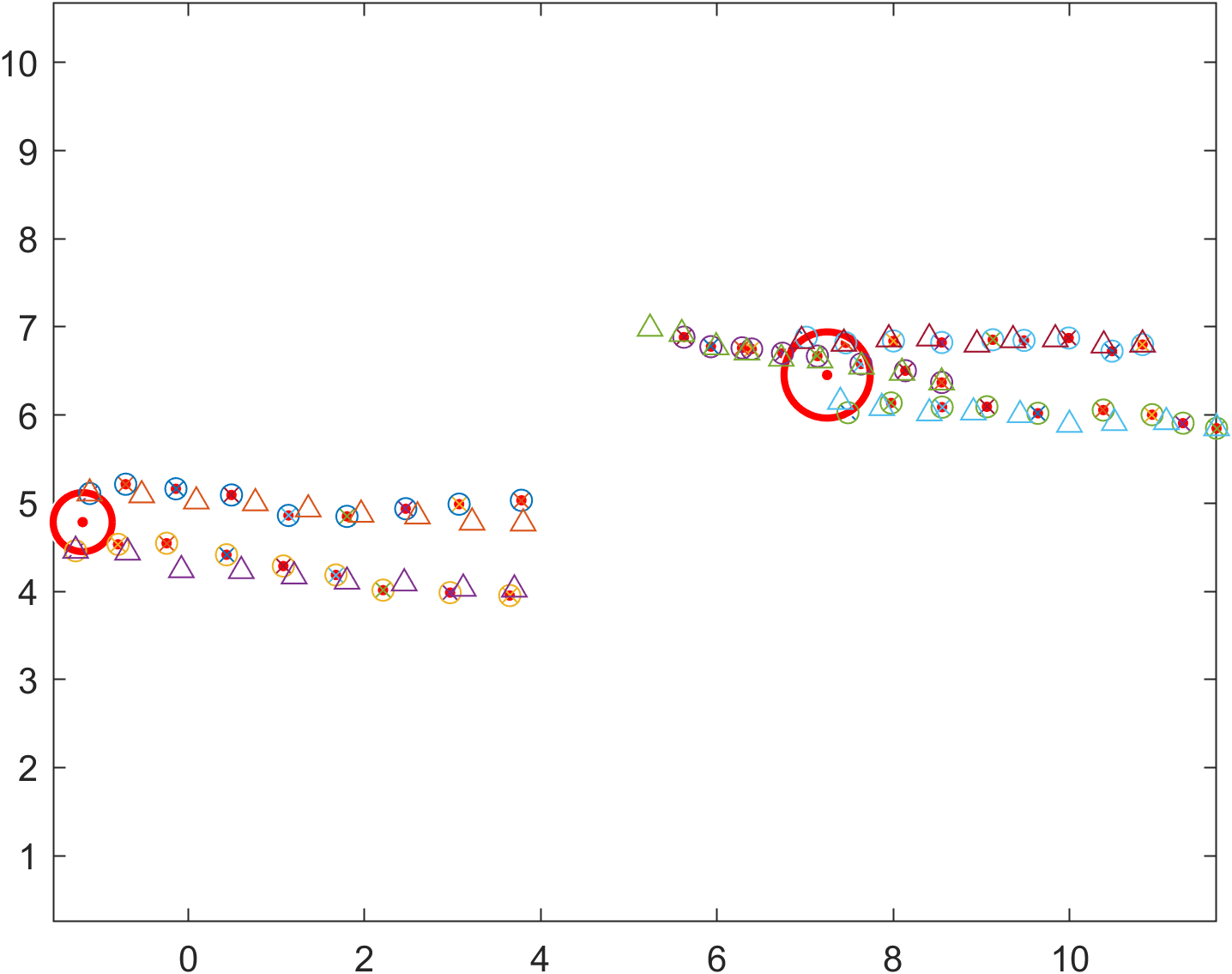}
         \caption{$\lambda_{1} = 0.7, \lambda_{2} = 0.3$}
         \label{fig:t4c4}
     \end{subfigure}
     \hfill
     \begin{subfigure}[b]{0.23\textwidth}
         \centering
         \includegraphics[width=\textwidth]{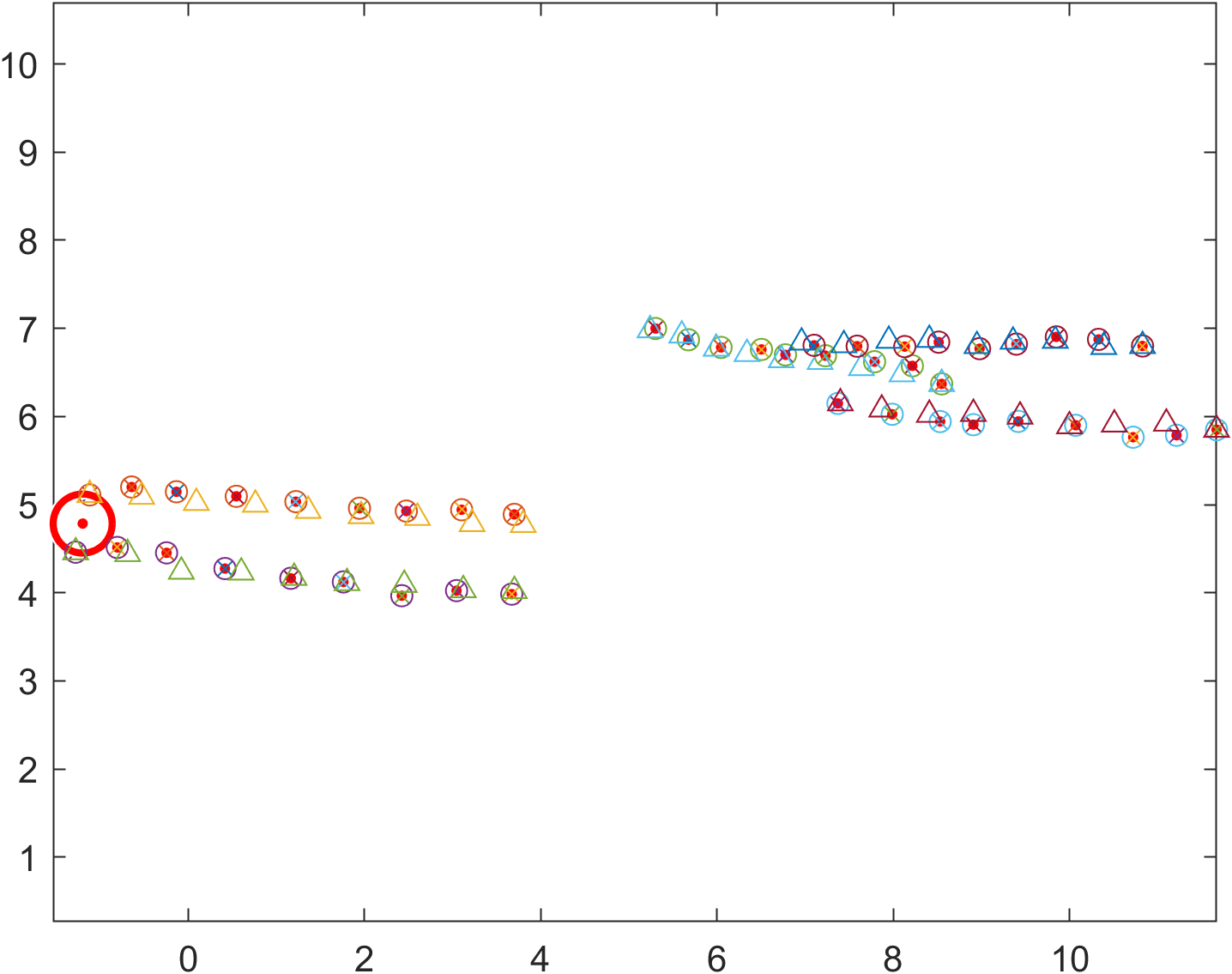}
         \caption{$\lambda_{1} = 1, \lambda_{2} = 0$}
         \label{fig:t4c5}
     \end{subfigure}
        \caption{Effect of varying cost parameters $\lambda$ on the clustering process for the Trajnet++ dataset. Different combinations of parameters are explored as $\lambda$ is varied from $0$ to $1$. The triangles represent the true data from Trajnet++ while the $\times$ represent the predicted data and the red ellipses indicate presence of a cluster between agents.}
        \label{fig:t4}
\end{figure}

Additionally, the effect of the parameters $\lambda_{1},\lambda_{2}$ are explored using a comparative study varying the parameter and noting the differences in the clusters formed for the same agent scenario. This explores the effect of both the goal based and the neighbour based cost criterion. The parameters are varied as $\lambda_{1} \in [0,1]$, $\lambda_{2} = 1 - \lambda_{1}$ and the resulting clusters formed over the simulation time are observed in Figure \ref{fig:t4}.

\subsubsection{Argoverse 2}
The Argoverse 2 dataset focuses on vehicular agents on various types of roadways but also includes pedestrian and cyclist agents as well. The results from applying the algorithm to the Argoverse 2 data is provided in this section. We focus on individual scenarios within the dataset to highlight the capabilities of the algorithm in different situations. 

\begin{figure}
     \centering
     \begin{subfigure}[b]{0.22\textwidth}
         \centering
         \includegraphics[width=\textwidth]{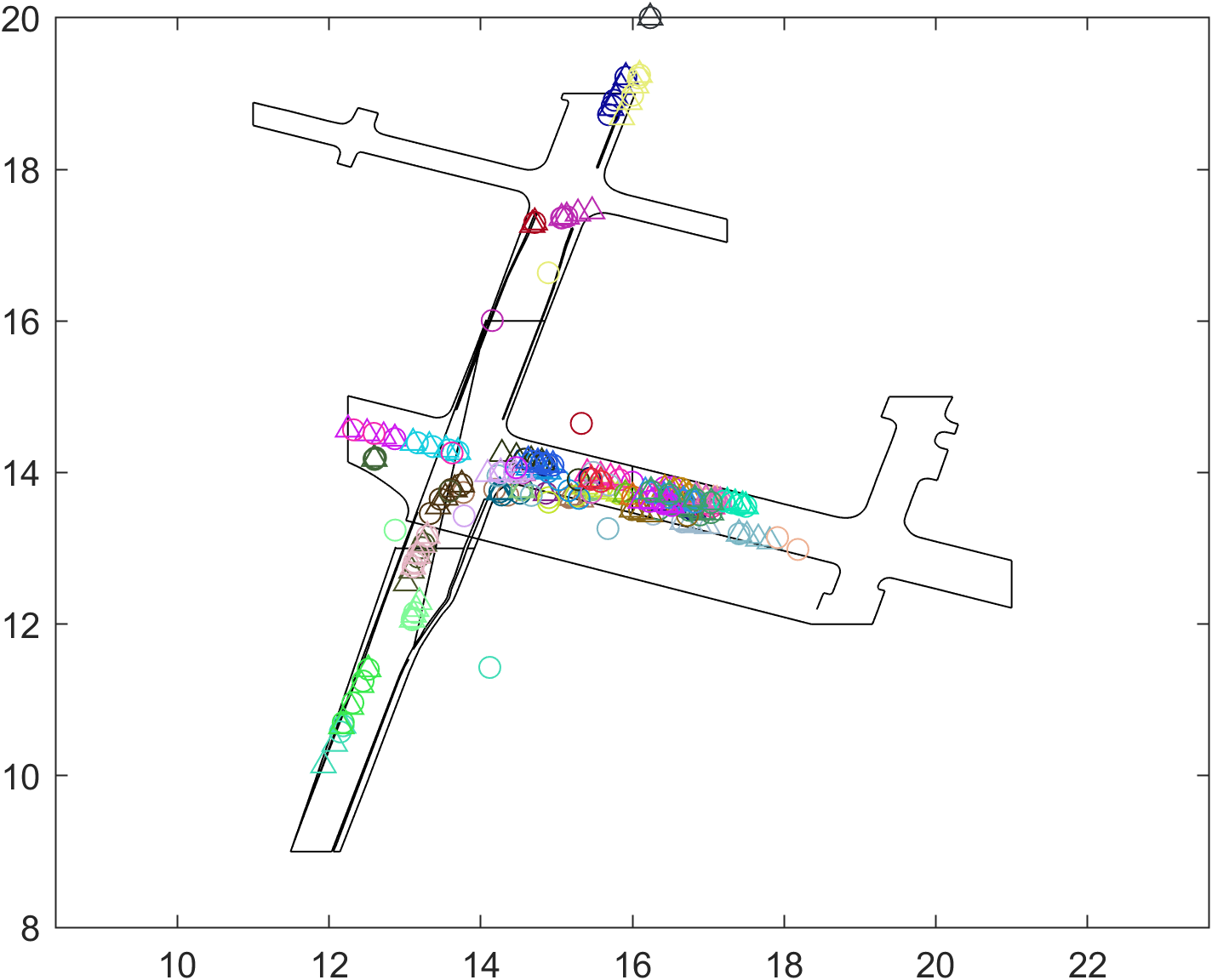}
         \caption{$t = 1-2s$}
         \label{fig:ar1t1}
     \end{subfigure}
     \hfill
     \begin{subfigure}[b]{0.22\textwidth}
         \centering
         \includegraphics[width=\textwidth]{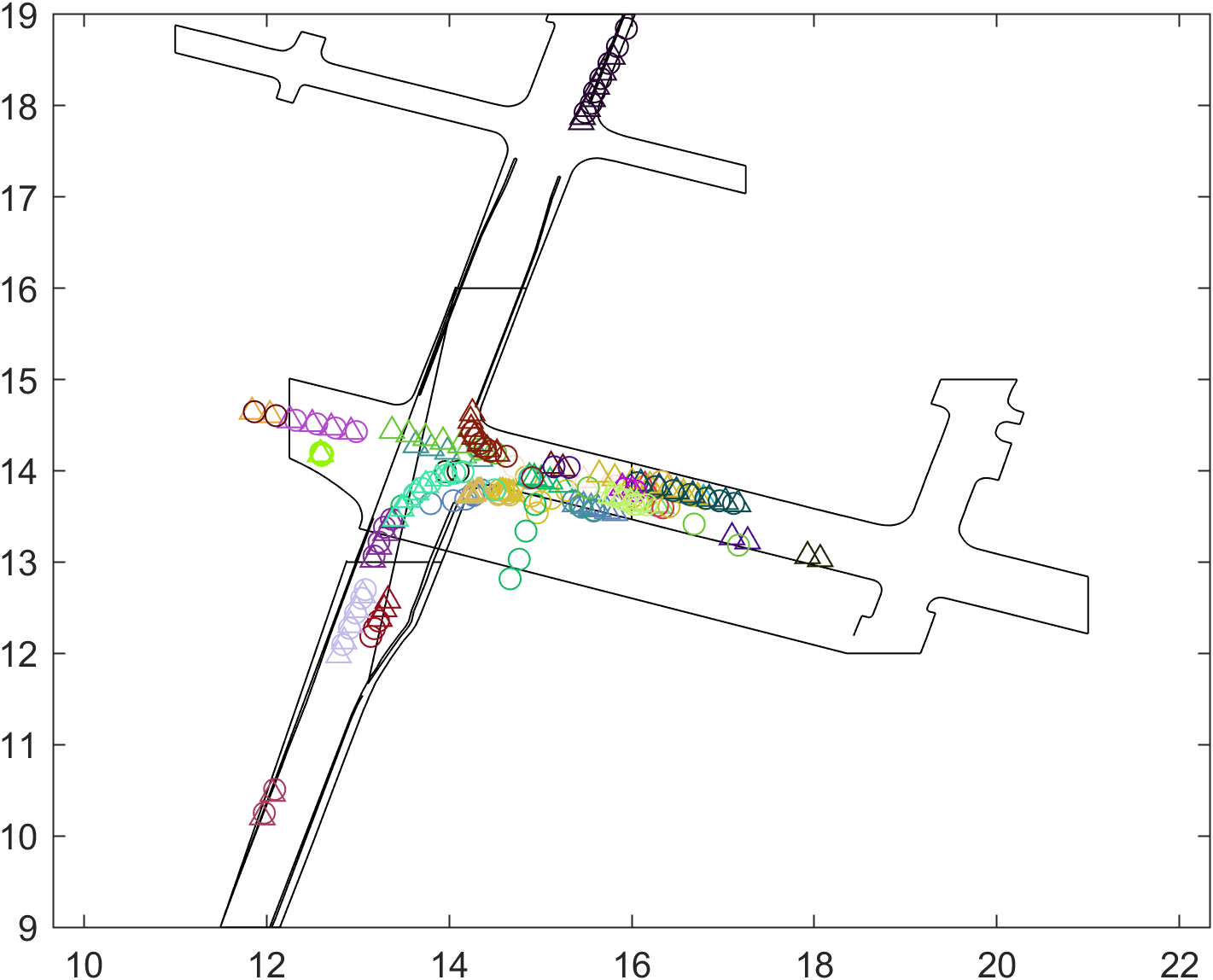}
         \caption{$t = 2-5s$}
         \label{fig:ar1t2}
     \end{subfigure}
     \hfill
     \begin{subfigure}[b]{0.22\textwidth}
         \centering
         \includegraphics[width=\textwidth]{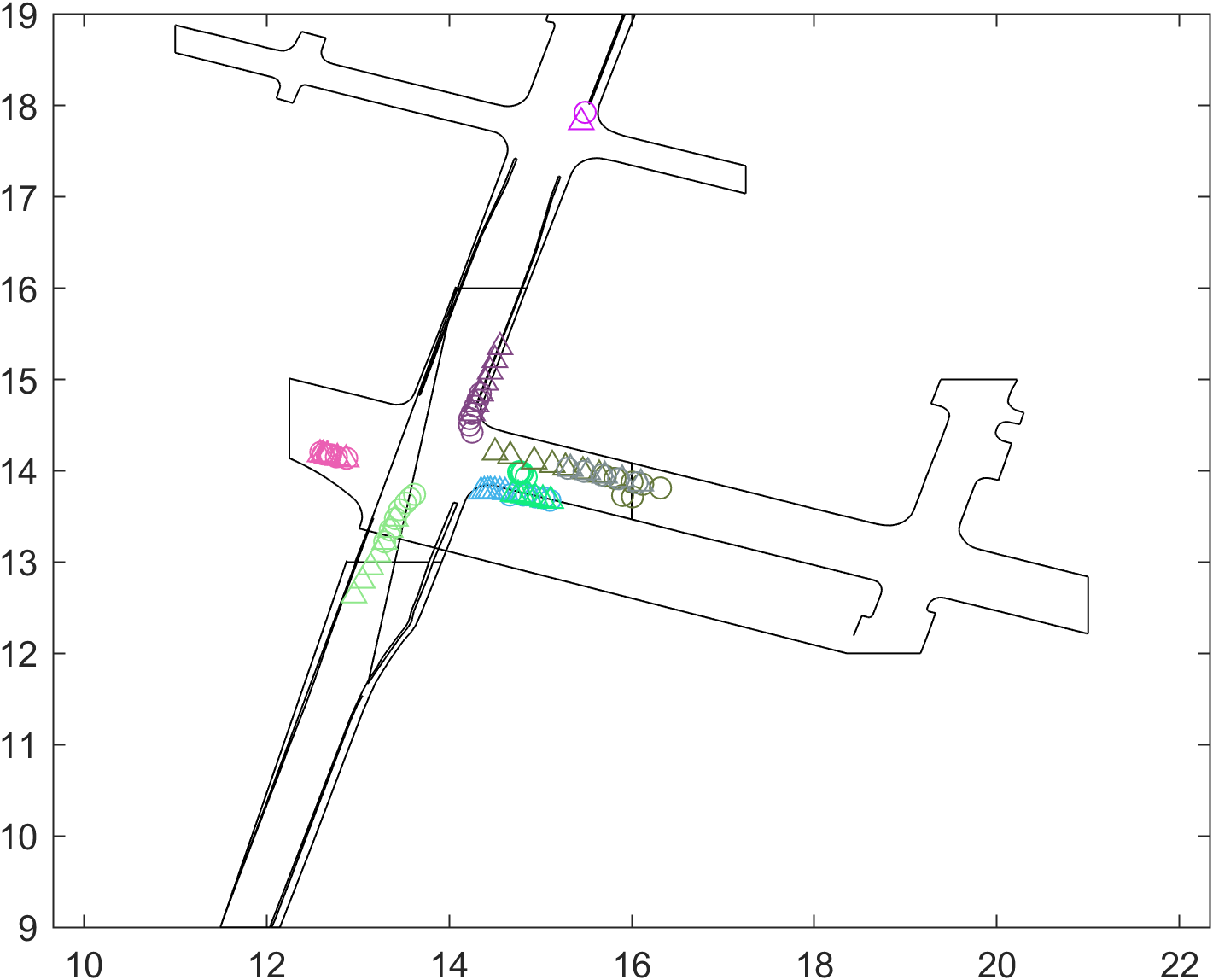}
         \caption{$t = 5-8s$}
         \label{fig:ar1t3}
     \end{subfigure}
     \hfill
     \begin{subfigure}[b]{0.22\textwidth}
         \centering
         \includegraphics[width=\textwidth]{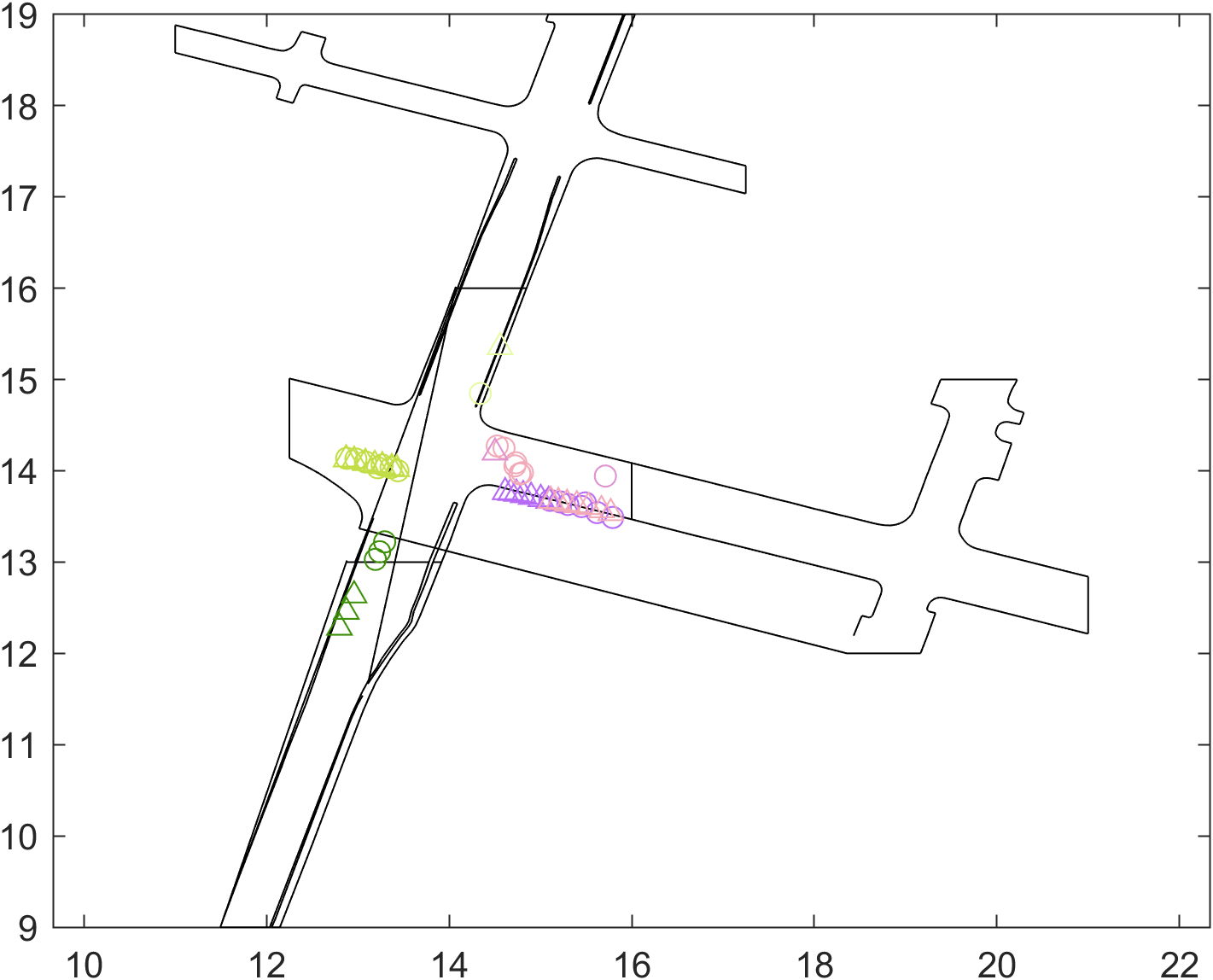}
         \caption{$t= 8-11s$}
         \label{fig:ar1t4}
     \end{subfigure}
        \caption{Agent trajectories over the course of time from $t = 1 - 11$s. The simulations have a timestep of 0.1s.}
        \label{fig:ar1t}
\end{figure}

\begin{figure}
    \centering
      \begin{subfigure}[b]{0.22\textwidth}
         \centering
         \includegraphics[width=\textwidth]{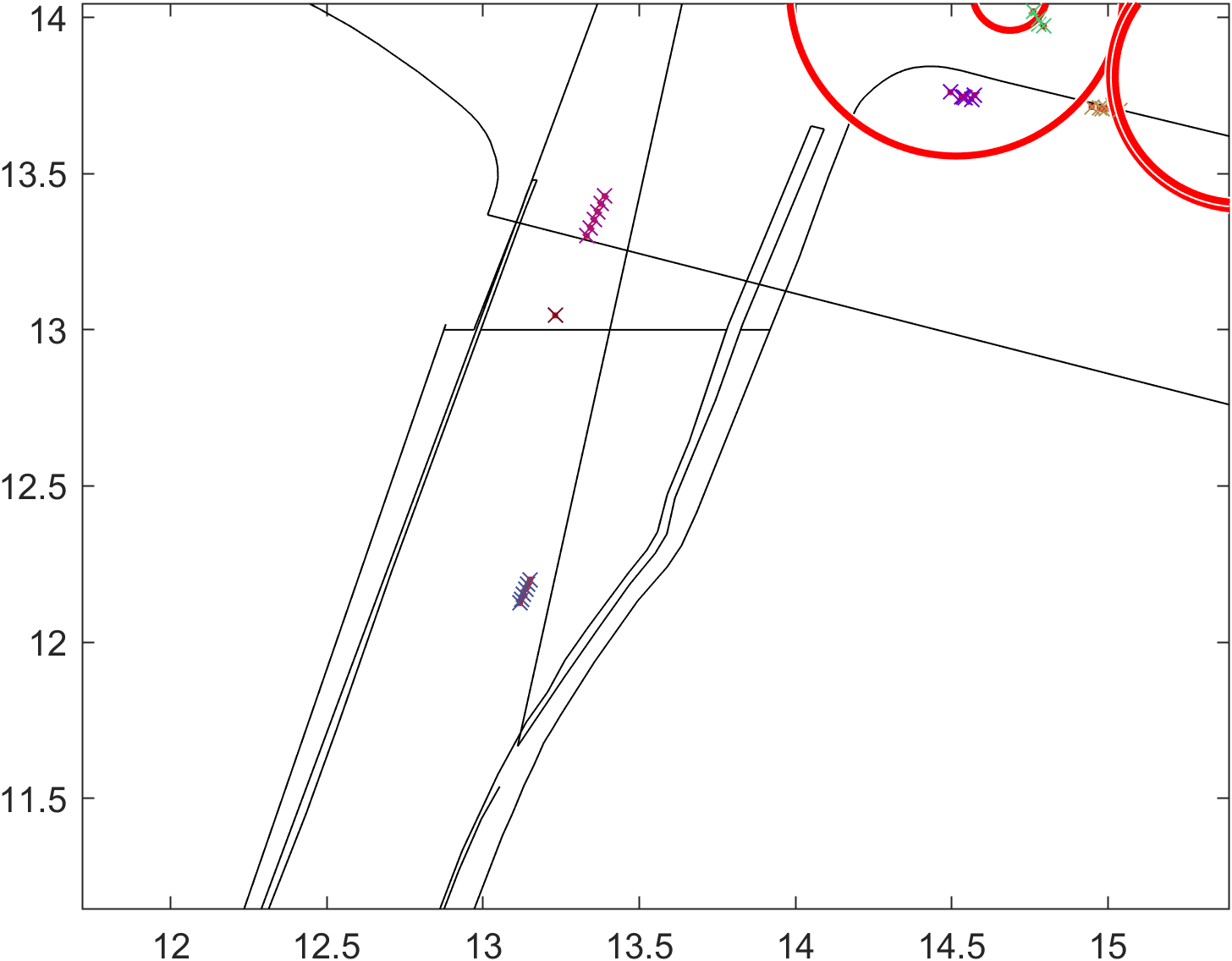}
         \caption{$t = 7.0-7.5s$}
         \label{fig:ar2t1}
     \end{subfigure}
     \hfill
           \begin{subfigure}[b]{0.22\textwidth}
         \centering
         \includegraphics[width=\textwidth]{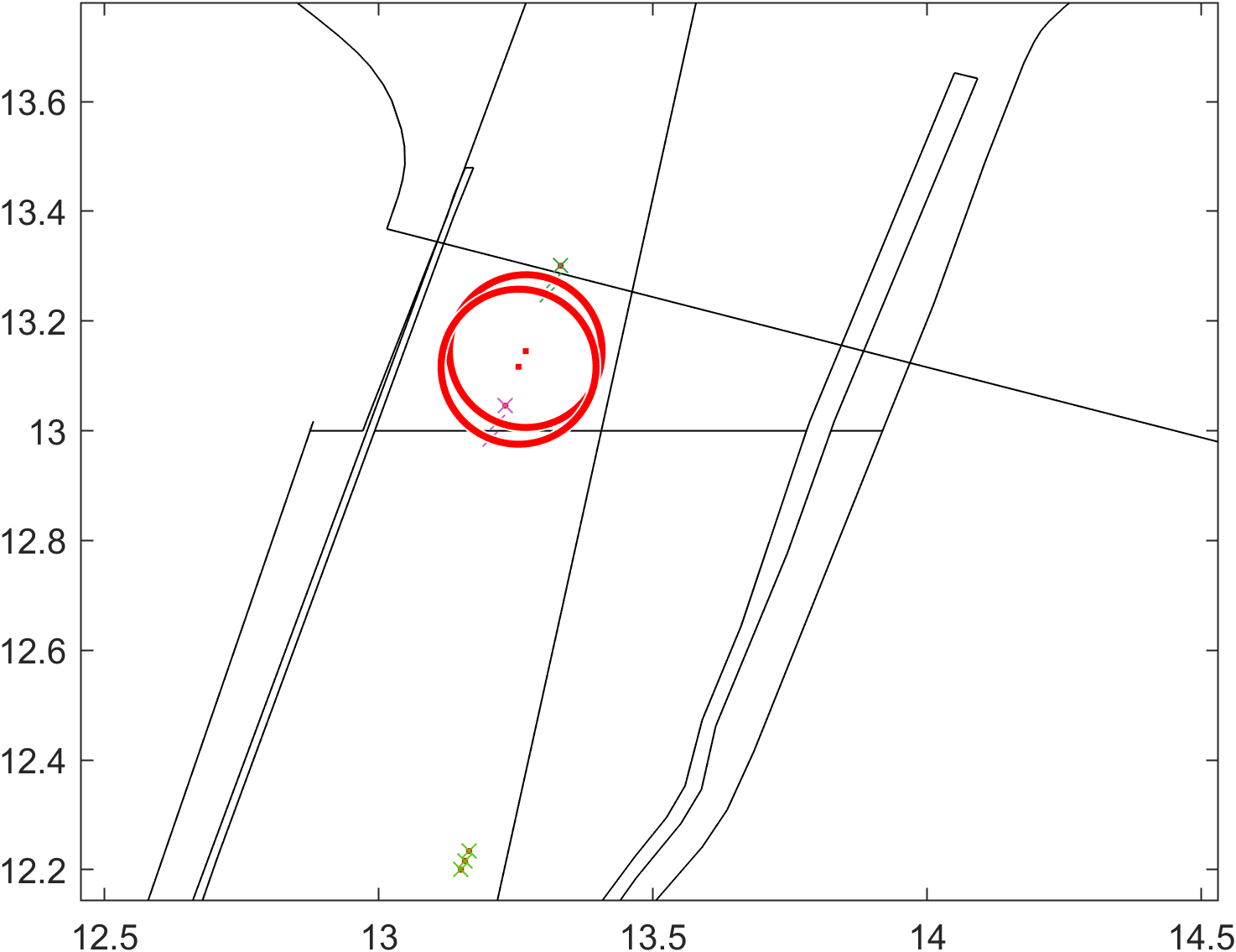}
         \caption{$t = 7.5-7.7s$}
         \label{fig:ar2t2}
     \end{subfigure}
     \hfill
       \begin{subfigure}[b]{0.22\textwidth}
         \centering
         \includegraphics[width=\textwidth]{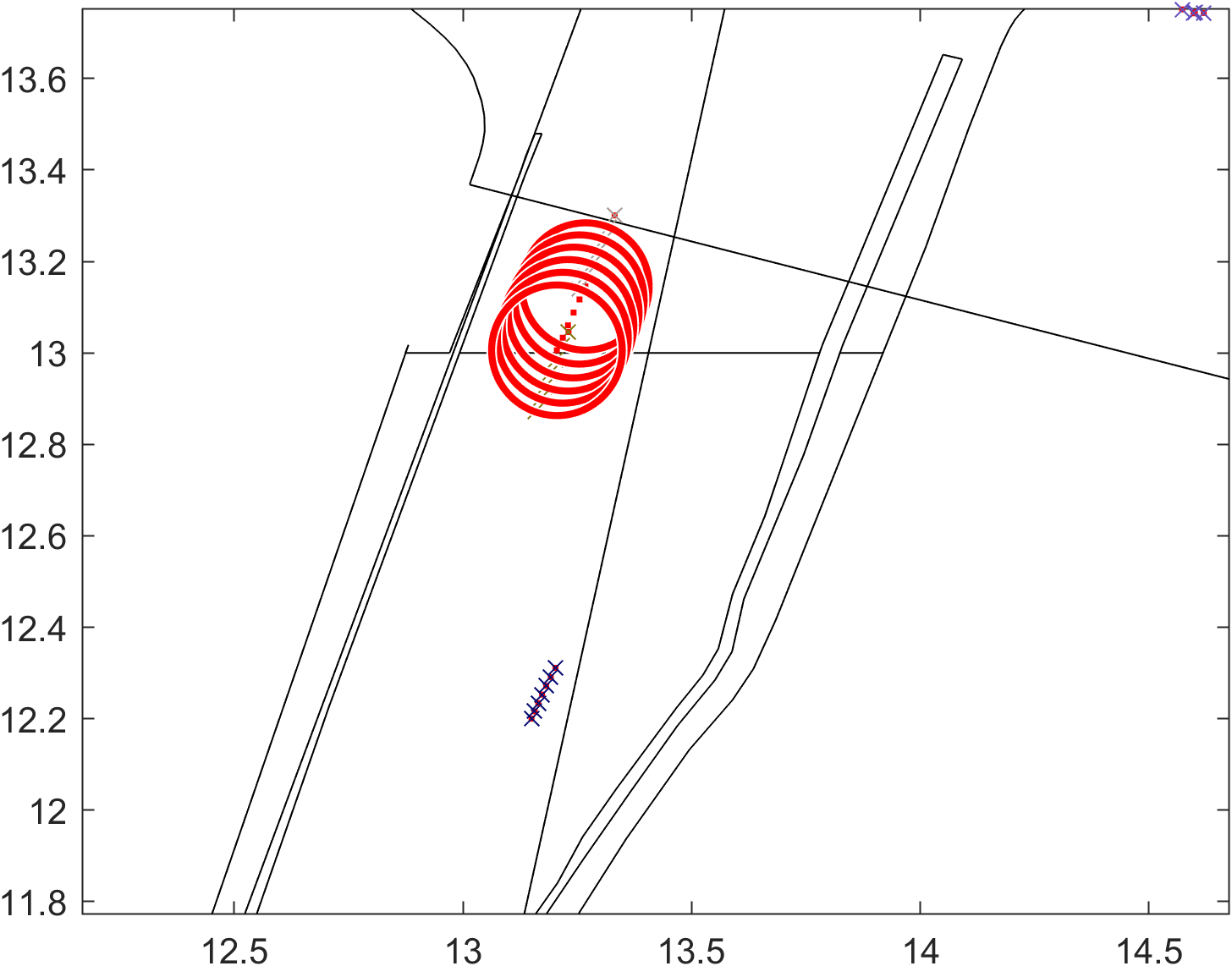}
         \caption{$t = 7.5-8.1s$}
         \label{fig:ar2t3}
     \end{subfigure}
    \caption{Close-up view of the trajectory of one cluster with two members, where the red circles centered on the cluster mean have a diameter which is equal to the distance between its members. represent the cluster centers (mean) over time and the radius of the red circles, encompassing the cluster members, represent the distance between the cluster mean and one of the agents. The cluster consists of two members going in the same direction along the road while another agent in close proximity is not included in the cluster as it is going in the opposite direction despite being in close enough proximity.}
    \label{fig:ar2}
\end{figure}

Figure \ref{fig:ar1t} illustrates a scenario of a roadway intersection with vehicles driving along the road while other vehicles wait at the intersection. The clustering algorithm prioritizes grouping agents traveling in the same direction within a certain distance. This distance depends on user-derived variable but can be adjusted for different general speeds and density of the road. Figure \ref{fig:ar2} depicts a close up view of the progression of a cluster with two members over time, where the red circle is centered at the cluster mean and has a diameter equal to the distance between the agents. From Figures \ref{fig:ar1t} and \ref{fig:ar2}, it is clear that the clustering algorithm prioritizes grouping agents that are traveling in the same direction. The range of the grouped agents does depend on user defined parameters and potentially would be adjusted for different scenarios depending on the location of the concerned agent in the scene, the type of roadway, the density of users on the road, the distance to the final goal etc. 

\begin{figure}
     \centering
     \begin{subfigure}[b]{0.22\textwidth}
         \centering
         \includegraphics[width=\textwidth]{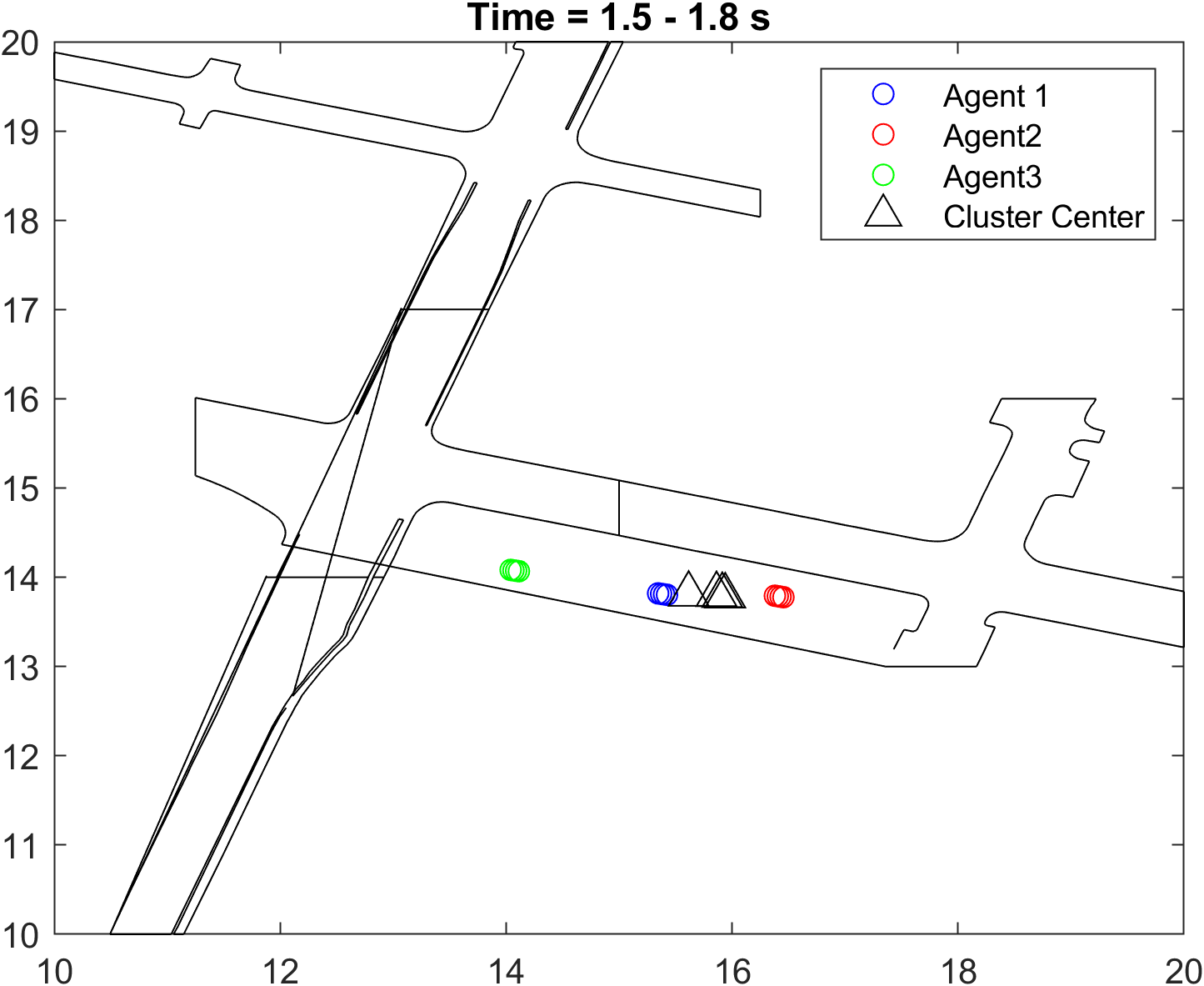}
         \caption{$t = 1.5s - 1.8s$}
         \label{fig:s1c1}
     \end{subfigure}
     \hfill
     \begin{subfigure}[b]{0.22\textwidth}
         \centering
         \includegraphics[width=\textwidth]{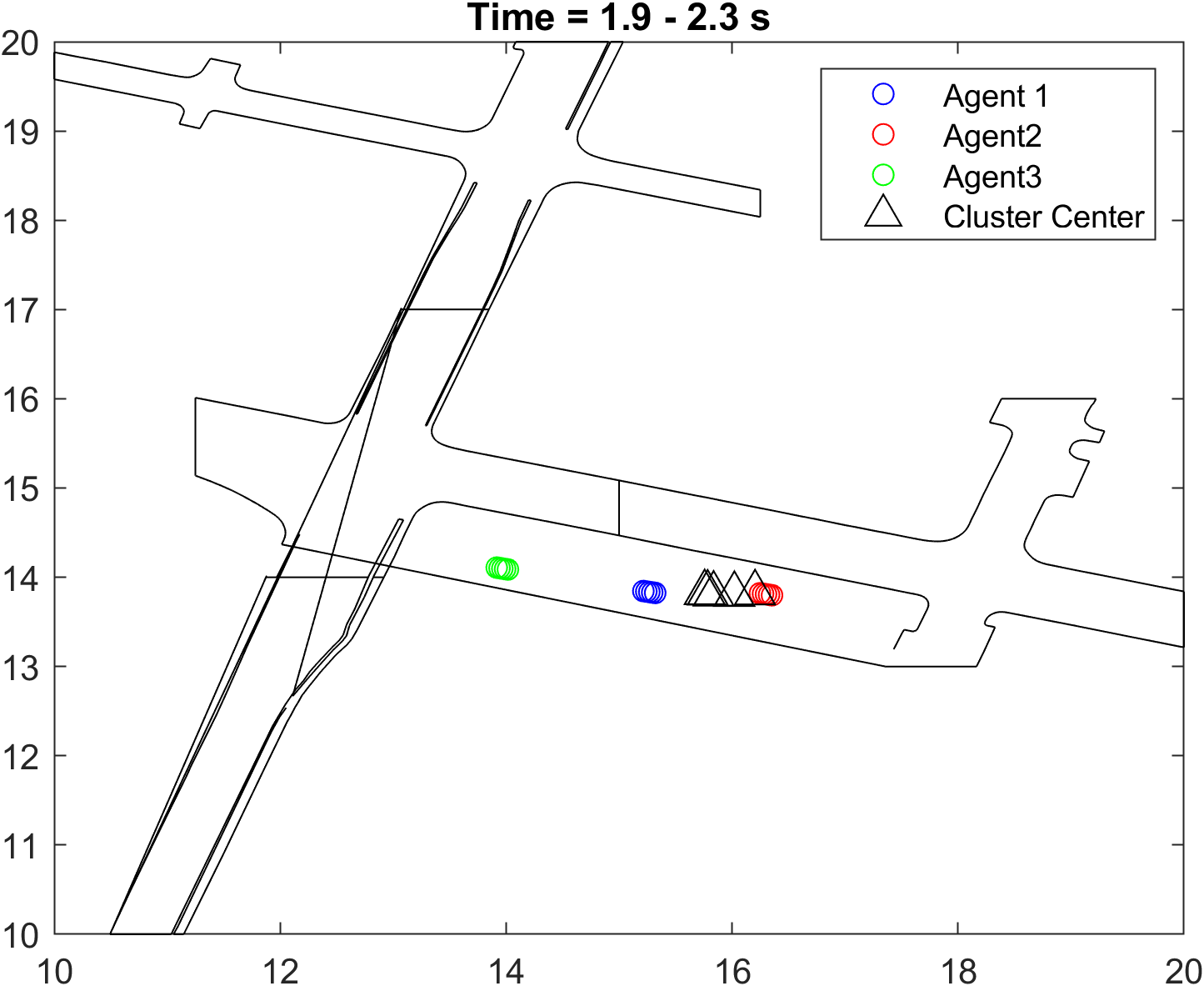}
         \caption{$t = 1.9s - 2.3s$}
         \label{fig:s2c1}
     \end{subfigure}
     \hfill
     \begin{subfigure}[b]{0.22\textwidth}
         \centering
         \includegraphics[width=\textwidth]{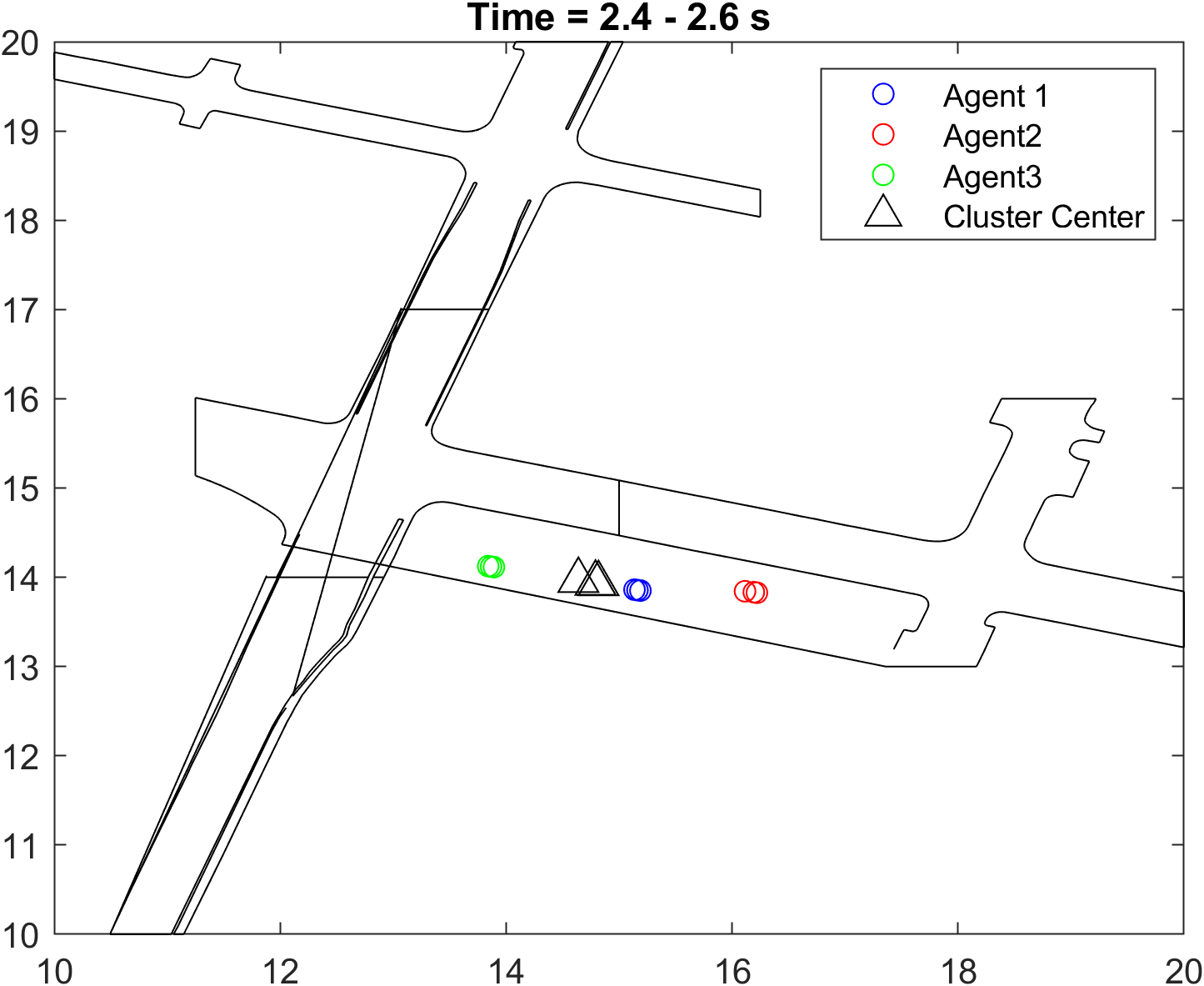}
         \caption{$t = 2.4s - 2.6s$}
         \label{fig:s3c1}
     \end{subfigure}
     \hfill
     \begin{subfigure}[b]{0.22\textwidth}
         \centering
         \includegraphics[width=\textwidth]{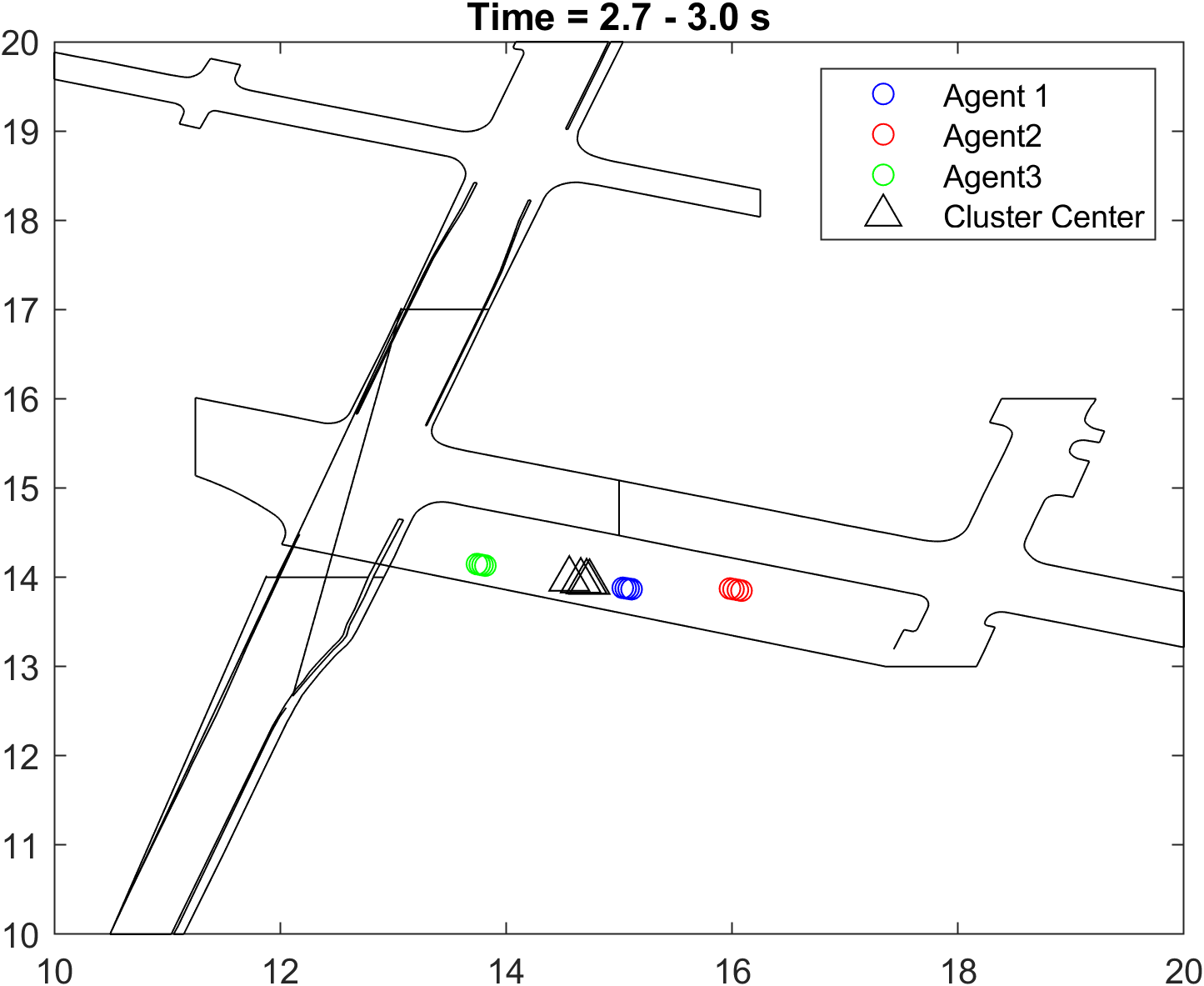}
         \caption{$t = 2.7s - 3.0s$}
         \label{fig:s4c1}
     \end{subfigure}
        \caption{Enlarged and isolated view of position prediction of dynamic cluster associated with a single agent (Agent $1$ in blue). The agent is initially clustered with Agent $2$ in red and later clustered with Agent $3$ in green. The cluster center which is the mean position of the cluster members is represented as a black triangle.}
        \label{fig:ar3}
\end{figure}

In Figure \ref{fig:ar3}, the effects of the dynamic clustering, are visualized. Agent 1 (blue) is initially close to Agent 2 (red) and were grouped into the same cluster. At a future time, Agent 1 has passed Agent 2 and is closer to Agent 3 (green), so the clustering algorithm has split the initial cluster and re-clustered Agent 1 and 3 since they are closer in Euclidean distance and speed. The cluster centers (black triangles) are the mean position of the state estimate of all the individual members of the cluster. As a result, the cluster center skews towards its cluster members rather than a straight line that mimics Agent 1.

\section{CONCLUSIONS}
This paper has presented a novel method for the future motion prediction of a group of agents by incorporating agent grouping according to a cost based criterion combined with geometric distance. Agents are grouped according to their feature similarity and known intent using a hierarchical agglomerative clustering method, enabling the creation of clusters independent of size and number constraints. 
This work includes a new clustering method that accounts for the features of the agents without explicitly requiring several thresholds for the distinction between agent states. The proposed algorithm is able to incorporate various features of the agent interactions using only the information of its position and velocity. The automatic split and merge of clusters is also introduced here using the predictions at each time. Future work will conduct extensive simulations on large datasets with diverse scenarios to show that the algorithm is able to effectively cluster and accurately predict the agent trajectories.

\section*{ACKNOWLEDGMENT}
This research has been supported in part by NSF award ECCS-1924790.

\bibliographystyle{ieeetr}
\bibliography{main}

\end{document}